\documentclass[11pt,aps,preprint]{revtex4}
\usepackage{latexsym}
\usepackage{amsmath}
\usepackage{graphicx}
\usepackage{subfigure}

\makeatletter
\renewcommand*{\email}[1][]{\begingroup\sanitize@url\@email{#1}}
\makeatother

\begin{document}
\title{An UV Completion of Five Dimensional Scalar QED and Lorentz Symmetry}

\author{F. Marques} 
\email{fabriciomarques@if.usp.br}


\author{M. Gomes}
\email{mgomes@if.usp.br}

\author{A. J. da Silva}
\email{ajsilva@if.usp.br}

\affiliation{Instituto de F\'\i sica, Universidade de S\~ao Paulo\\
Caixa Postal 66318, 05315-970, S\~ao Paulo, SP, Brazil}%

\begin{abstract}
We study a five dimensional Horava-Lifshitz like scalar QED with dynamical exponent $z=2$. Consistency of  the renormalization procedure requires 
 the presence of  four quartic  and one six-fold scalar couplings   besides the  terms  bilinear in the scalar fields. We compute one-loop radiative corrections to the parameters in the original Lagrangian, employing dimensional regularization in the spatial part of the Feynman integrals and prove the relevant Ward identities. By using renormalization group methods,  we determine the behavior of the  coupling constants with changes in the energy and discuss the emergence of Lorentz symmetry at low energies. 
\end{abstract}  

\maketitle

\section{Introduction}
The use of Lagrangians exhibiting space-time anisotropy and equipped with high spatial derivative terms, Horava-Lifshitz (HL) like  models \cite{Hor1,Lif}, has attracted considerable attention in  the recent years. This is so because they allow for an ultraviolet completion of otherwise  nonrenormalizable models and in particular may lead to a consistent quantum gravity theory \cite{Hor1,Hor2}. It should be noticed that originally high spatial derivatives were  used  in the description of Lifshitz points in statistical mechanics studies \cite{Lif}. Further applications to statistical mechanics and condensed matter may be found in \cite{Fradkin,Gomes2}.
   
Considerable amount of work has been devoted
to study different facets of HL models. These studies encompass  quantum gravitational issues,  as black holes \cite{Cai}, renormalization features \cite{Barvinsky} and other aspects \cite{Lu}. Besides that, many studies have also been dedicated to non gravitational models \cite{Ferrari}.  In particular for scalar models, renormalization  aspects  have been treated  in \cite{Anselmi,Iengo1},    gauge theories similar to QED were studied in \cite{Iengo2,Gomes},  Ward identities and anomalies were considered in \cite{Gomes1,Arav, Adam}.

The   basic assumption behind these proposals is that, asymptotically, the equations of motion  are invariant under the rescaling $x^{i}\to b x^{i}, t\to b^z t$ where $z$, the so called dynamical critical exponent, is related with the ultraviolet behavior of the models. As space time anisotropy   breaks Lorentz symmetry,  to physically validate HL models at the low energy scale of the present Universe, it is necessary to prove that Lorentz invariance
is at least approximately realized  at small energies. Renormalization group arguments indicate that to achieve this behavior  it is required that the effective coefficients of the high  derivative terms in the Lagrangian should monotonically decrease as the energy decreases.

 In the last two decades,  models in more than four dimensions have aroused a great deal of  interest (see \cite{Shifman} and references therein). The reason is that compactification of extra dimensions  introduces  new scales and  new physics in  the desert separating the electroweak unification scale  ($10^2$ Gev) from  the Planck scale ($10^{19}$ Gev)  of the quantization of gravitation, the hierarchy problem. However, usual quantum field models  ($z=1$) are in general nonrenormalizable in more than four dimensions. 
 This work is dedicated to the study of $z=2$  scalar quantum electrodynamics in five dimensions, the highest dimension where this model is renormalizable. Actually, the model  is super renormalizable or nonrenormalizable  for dimensions lower or higher  than five, respectively. 

We would like to point out to some earlier studies related to this subject.  Reference \cite{Gomes}   provided a study of Lorentz symmetry restoration and
a discussion of anomalies in a four dimensional HL like spinor and scalar QED.  That work was followed by \cite{Gomes3} in which, also in four dimensions,  the anomalous magnetic momentum was determined  and a complete one-loop renormalization analysis was presented. In five dimensions we are aware
of the work \cite{Iengo2} on spinor HL like QED showing that a great simplification occurs at very high energies where the usual spatial terms, i.e.,  linear terms in the spatial derivatives,
may be neglected; in particular, the gauge coupling constant is not renormalized. This simplicity was also pursued in \cite{Gomes} and \cite{Gomes3} the usual term being also absent. Differently, we consider here the dynamics of the more general renormalizable scalar model obeying gauge symmetry and charge conjugation. The presence of  usual terms, quadratic in the spatial derivatives, turn unfeseable the complete calculation of the Green functions.  In spite of this,  it is still possible to obtain one-loop renormalization constants which allow for the determination of relevant 
renormalization group $\beta $ functions. Using these results, we analyze the evolution of the parameters of the theory and determine a range for which Lorentz symmetry may be restored.

 One possible usefulness of this work is the following. Five dimensional scalar QED with $z=1$ is nonrenormalizable; in this situation one may still use it as an effective theory for small energies up to some scale $\Lambda$. To fix $\Lambda$ we may consider the Lagrangian with $z=2$ which is renormalizable but breaks Lorentz symmetry. However, if we could find, for small energies, a range of values for which the Lorentz symmetry is approximately realized, we may take these energies as the ones where the effective theory with $z=1$ is approximately correct.

This work is organized as follows. In section \ref{section2} we  introduce the model, state the Feynman rules needed to compute the radiative corrections, present the degree of superficial divergence and  show the results for the  one-loop  vertex functions. Explicit calculations of the divergences and renormalization are provided in the Appendix, where we also verified the Ward identites obeyed by the vertex functions.
 In section \ref{section4}, by using renormalization group methods, the relevant $\beta$ functions are computed. Finally, in section \ref{section5} we present a summary and the conclusions of this work. 
\section{The model}\label{section2}
In this work we study a $z=2$ version of five dimensional scalar QED described by the Lagrangian density
\begin{eqnarray}
  \begin{aligned}
  \cal {L} =& \frac{1}{2}F_{0 i} F_{0 i} 
                 - \frac{a_1^2}{4} F_{i j} F_{i j} 
                 - \frac{a_2^2}{4} \partial_l F_{i j} \partial_l F_{i j} \\
               & + ( D_{0} \phi)^{*} D_{0} \phi 
                 - b_1^2 ( D_{i} \phi)^{*} D_{i} \phi
                 - b_2^2 ( D_{i} D_{j} \phi)^{*} D_{i} D_{j} \phi
                 - m^2  \phi^{*} \phi  \\
               & -i e b_3^2 F_{ij} ( D_{i} \phi)^{*} D_{j} \phi
               - \frac{e^2}{2} b_4^2 F_{ij}F_{ij} \phi^{*} \phi,
              \end{aligned}
  \label{Lagrangian}
\end{eqnarray}
where $D_{0,i}=\partial_{0,i}-i e A_{0,i}$ is the gauge covariant derivative. The parameters $a_{2}$ and $b_{i}$ with $i=2,3,4$, which control the high derivative terms, are taken to be dimensionless in  momentum units. From that and taking into account the   dimension  six of ${\cal L}$, 
we get that the dimensions of $\phi$ and $A_{i}$ are equal  to one whereas the dimension of $A_{0}$ is two. The parameters $a_{1}^{2}, b_{1}^{2 }$ and $m$ have dimension two and $e$ is dimensionless. The above expression is the most general gauge invariant  Lagrangian containing at most two scalar fields.  Integrating by parts, other possible terms, as for example $\partial_{j}F_{ij}\partial_{l}F_{il} $, may be reduced to the ones in (\ref{Lagrangian}) .

We choose to work in a strict Coulomb gauge by adding to (\ref{Lagrangian}) the gauge fixing Lagrangian
\begin{equation}
  {\cal L}_{\text{GF}} = \frac{\eta}{2} ( \partial_{i} A_{i})^2
  \label{CoulombGaugeFixingLagrangian}
\end{equation}
and letting $\eta$ tend to infinity. Notice that gauge invariance and charge conjugation ($\phi \leftrightarrow \phi^{*}$ and $A_{\mu}\rightarrow -A_{\mu}$)  forbid the appearance  of pure gauge monomials, without
scalar field factors and containing more than two gauge fields.  However, we will show shortly that terms with four and six scalar fields have to be included.
Using the above Lagrangian, we  obtain the propagators and interacting vertices,

1. For the gauge field:
\begin{equation}
 \langle T A_i(k) A_j(-k)\rangle =i\frac{\delta_{ij}-\frac{k_{i}k_{j}}{{\vec k}^{2}}}{k^2_0-a_{1}^{2}\vec{k}^{2}-a_{2}^{ 2}\vec{k}^4+i\epsilon};\quad \quad\langle T A_0(k) A_0(-k)\rangle =\frac{i}{\vec{k}^{2}}
\end{equation}
and $\langle T A_{0}(k)A_{i}(-k)\rangle=0$.

2. For the scalar field:
\begin{equation}
\langle T \phi(k) \phi ^{*}(-k)\rangle= \frac{i}{k_0^2 - b_1^2 \vec{k}^2 - b_2^2 \vec{k}^4 - m^2+i\epsilon}.
\end{equation} 

There are four 3-linear vertices which we 
 label  as $V_{3X}$, $X=A,B,C,D$. By taking the Fourier transforms 
of these interaction terms and taking the momenta always entering at the vertex, one finds 
their expressions in momenta space to be:
\begin{eqnarray}
    V_{3A}(p,k,k') &=& e A_0(p) \phi(k) \phi^{*}(k') \times ( p_0 + 2 k_0 ), \\
    V_{3B}(p,k,k') &=& - e b_1^2 A_i(p) \phi(k) \phi^{*}(k') \times ( p_i + 2 k_i ), \\
    V_{3C}(p,k,k') &=& - e b_2^2 A_i(p) \phi(k) \phi^{*}(k') \times ( p_j + 2 k_ j )
     \big\{ ( p_i + k_i ) (p_j + k_j ) + k_i k_j \big\} ,\\
    V_{3D}(p,k,k') &=& - e b_3^2 A_i(p) \phi(k) \phi^{*}(k') \times
    \left\{ 
    k_i(\vec{k'}^{\,2} + \vec{k'}\cdot \vec{k}) 
    - k'_i(\vec{k}^{\,2} + \vec{k'}\cdot \vec{k})
    \right\},
    \label{V3Expressions}
\end{eqnarray}
where $k'=-k-p$. There are also  five 4-linear vertices:
\begin{eqnarray}
V_{4A}(p,p',k,k') &=& e^2 A_0(p) A_0(p') \phi(k) \phi^{*}(k'),\\
    V_{4B}(p,p',k,k') &=& - e^2 b_1^2 A_i(p) A_i(p') \phi(k) \phi^{*}(k'), \\
    V_{4C}(p,p',k,k') &= &- e^2 b_2^2 A_i(p) A_j(p') \phi(k) \phi^{*}(k') \times
     \big\{ k_i k_j + k'_i k'_j - k'_i k_j - k_i k'_j\nonumber \\ 
     &&
     - 2 \vec{k}\cdot \vec{k'} \delta_{ij} 
     - \vec{p}\cdot \vec{p'} \delta_{ij} 
     - p'_i (k_j + k'_j) 
     - \vec{p'} \cdot (\vec{k} + \vec{k'}) \delta_{ij} \big\}, \\
    V_{4D}(p,p',k,k') &=& - e^2 b_3^2 A_i(p) A_j(p') \phi(k) \phi^{*}(k') \times
    \left\{  
    p'_i(k_j + k'_j) - \delta_{ij} \vec{p'} \cdot (\vec{k}+\vec{k'})
    \right\}  \\
    V_{4E}(p,p',k,k') &=& - e^2 b_4^2 A_i(p) A_j(p') \phi(k) \phi^{*}(k') \times
    \left\{ p'_i p_j - \delta_{ij} \vec{p'} \cdot \vec{p} \right\},
    \label{V4Expressions}
\end{eqnarray}
where the momenta satisfy $k' = -k-p-p'$. There is also a vertex with five fields given by
\begin{equation}
    V_{5}(p_1,p_2,p_3,k,k') = - 2 e^3 b_2^2 A_i(p_1) A_i(p_2) A_j(p_3) \phi(k) \phi^{*}(k') \times
    \big( k_j - k'_j \big) ,  \label{V5Expressions}
\end{equation}
where $p_1+p_2+p_3+k+k'=0$ and  a vertex with six fields,
 \begin{equation}
 V_{6}(p_{1},p_{2},p_{3},p_{4},k,k')=e^{4}A_{i}(p_{1}) A_{i}(p_{2})A_{j}(p_{3}) A_{j}(p_{4}) \phi(k) \phi^{*}(k'),
 \end{equation}
with the momenta satisfying $\sum_{i=1}^{4} p_{i}+k+k'=0$.

 By using these expressions,   we may compute  the degree of superficial divergence for a generic Feynman diagram $\gamma$
\begin{equation}
\delta(\gamma)=6- N_{\phi}-N_{A_{i}}-2N_{A_{0}}-2\nu_{3B}-2\nu_{4B},\label{1}
\end{equation}
where $N_{\cal O}$ denotes the number of external lines of the field ${\cal O}$ and $\nu_{\cal O}$ is the number of vertices of the type $V_{\cal O}$ in $\gamma$. Notice from (\ref{1})  that graphs without external gauge field lines but either with four or six external scalar lines are quadratically and logarithmically divergent, respectively. Therefore, for consistency of the renormalization process, one should enlarge our model and add to (\ref{Lagrangian}) the terms given by
\begin{eqnarray}
{\cal L}_{\phi}&=&
    \xi_1 \left[ \phi^{*} (D_i D_i \phi) + (D_i D_i \phi)^{*} \phi \right] \phi^{*} \phi\nonumber
    \\ &&
    + \xi_2 \left[\phi^{*}(D_i \phi)\phi^{*}(D_i \phi)+(D_i\phi)^{*}\phi(D_i \phi)^{*}\phi \right]\nonumber
    \\ &&
    + \xi_3 \phi^{*}(D_i \phi)(D_i \phi)^{*}\phi- \frac{ \lambda}{4} (\phi^{*} \phi)^2 - \frac{g}{6}(\phi^{*} \phi)^3.\label{Lagrangian1}
\end{eqnarray}
Notice that, unless for the term with $\lambda$, all these vertices have ultraviolet dimension $6$ and are therefore renormalizables; they do not modify the power counting given in (\ref{1}).  The vertex with the coupling $\lambda$ has operator dimension four, it is super renormalizable and modify the power counting; a term $-2 \nu_{\lambda}$ has to be added to the rhs of (\ref{1}).  To keep the ultraviolet divergences under control, the spatial part of the Feynman integrals will be regularized to $d=4-\epsilon$ dimensions. 
It is also convenient to introduce a parameter $\mu$ with momentum dimension two and make the following replacements:
\begin{equation}
  e \rightarrow e\mu^{\epsilon/4} 
  \, , \quad
  \lambda \rightarrow \lambda \mu^{\epsilon/2}
  \, , \quad
  g \rightarrow g \mu^{\epsilon} 
  \, , \quad
  \xi_n \rightarrow \xi_{n} \mu^{\epsilon/2},
  \qquad  \qquad
  \label{DimensionlessCouplingsz2d4}
\end{equation}
with $n=1,2,3$. After the pole part of the integrals
have been removed, we will let $\epsilon\to 0$.
In the Appendix we determined the counterterms needed to eliminate these would be divergences. Using those results, we obtain 
 the gauge field two  point vertex functions 
\begin{equation}
  \Gamma^{(2)}_{00}(p) =\left(1 + \frac{\alpha}{8}\ln{\mu} \right) \vec{p}^{\,2} +
   (\text{finite part}) \, ,
  \label{FinalRenormalized1PI2Points00}
\end{equation} 

\begin{equation}
  \Gamma^{(2)}_{0i}(p) = \Gamma^{(2)}_{i0} =  \left(1 + \frac{\alpha}{8}\ln{\mu} \right)p_0 p_i +
   (\text{finite part})
  \label{FinalRenormalized1PI2Points0i}
\end{equation}
and

\begin{eqnarray}
  \begin{aligned}
  \Gamma^{(2)}_{ij} (p)
  &
  = \left( 1 + \frac{\alpha}{8}\ln{\mu} \right) \delta_{ij}p_0^2
  -\left( a_1^2 - \frac{\alpha}{8} R \ln{\mu} \right)
  (\delta_{ij} \vec{p}^{\,2} - p_i p_j) 
  \\
  &
  - \left( a_2^2 - \frac{\alpha}{8} S \ln{\mu} \right)
      (\delta_{ij}\vec{p}^{\,2} - p_i p_j) \vec{p}^{\,2}+
   (\text{finite part}),
  \end{aligned}
  \label{FinalRenormalized1PI2Pointsij}
\end{eqnarray}
where $\alpha=\frac{e^{2}}{16\pi^{2}b_{2}}$ and $R$ and $S$ are defined in (\ref{RS}).

It should be stressed that the two point vertex function of the gauge field that we are considering is restricted to its transverse part as its longitudinal part is meaningless.

We have also
\begin{eqnarray}
  \begin{aligned}
    \Gamma^{(2)}(p)& = \left( 1 - \frac{\alpha}{2} \ln{\mu} \right) p_0^2
    -  \left( b_1^2 - \frac{Q_1}{2}  \ln{\mu} \right) \vec{p}^{\,2}
    -  \left( b_2^2 - \frac{Q_2}{2}  \ln{\mu}  \right) \vec{p}^{\,4} \\
    &- \left( m^2  - \frac{Q_3 }{2} \ln{\mu} \right)
    +(\text{finite part}),
  \end{aligned}
  \label{FinalRenormalized1PI2PointsMatter}
\end{eqnarray}
for the renormalized two point function of the scalar field, with $Q_{1}$, $Q_{2}$ and $Q_{3}$ given in (\ref{Q1}-\ref{Q3}),
\begin{eqnarray}
    \Gamma^{(3)}_{0}(p-p^{'}) = e \left( 1 - \frac{\alpha}{2} \ln{\mu} \right) (p_0 + p'_0)+
   (\text{finite part)},
  \label{FinalRenormalized1PI3Points0}
\end{eqnarray}
for the three point vertex function $\langle A_{0}\phi^{*}\phi\rangle$ and
\begin{eqnarray}
  \begin{aligned}
     \Gamma^{(3)}_{i}(p,-p^{'}) &=     
     e \left( b_1^2 - \frac{ Q_1}{2}  \ln{\mu} \right) (p_i + p'_i) \\
     &
     e  \left( b_2^2 - \frac{ Q_2}{2}  \ln{\mu}  \right) 
    \big( \, p_i (\vec{p}^{\,2} + \vec{p} \cdot \vec{p'} ) 
    + p'_i ( \vec{p'}^{\,2} + \vec{p} \cdot \vec{p'} ) \, \big) \\
    &
    e  \left( b_3^2 - \frac{ K_3 }{2} \ln{\mu} \right)
    \big( \, p_i ( \vec{p'}^{\,2} - \vec{p} \cdot \vec{p'}  ) 
    + p'_i ( \vec{p}^{\,2} - \vec{p} \cdot \vec{p'}  ) \, \big)  \\
    &
    + (\text{finite part}),
  \end{aligned}
  \label{FinalRenormalized1PI3Pointsi}
\end{eqnarray}
for the three point vertex function $\langle A^{i}\phi^{*}\phi\rangle$, with $K_{3}$ defined in (\ref{K3}).
As argued in the Appendix, these functions satisfy  the simplest Ward identities associated with current conservation.

 In the next section, we  employ these expressions to find  some of the $\beta$ functions of the model.

\section{Renormalization group and effective couplings}\label{section4}

We may now fix the renormalization group flows of the parameters of the model. The vertex functions $\Gamma^{(N_{A_{0}},N_{A_{i}},N_{\phi})}(p)$ satisfy the 't Hooft-Weinberg renormalization group equation

\begin{eqnarray}
  \begin{aligned}
  &\bigg[ 
  \mu \frac{ \partial}{ \partial \mu}  
  + a_1 \beta_{a_1} \frac{ \partial}{ \partial a_1} 
  + \beta_{a_2} \frac{ \partial}{ \partial a_2} 
  + b_1 \beta_{b_1} \frac{ \partial}{ \partial b_1} 
  + \beta_{b_2} \frac{ \partial}{ \partial b_2}
  + \beta_{b_3} \frac{ \partial}{ \partial b_3}
  + \beta_{b_4} \frac{ \partial}{ \partial b_4} + \beta_{e} \frac{ \partial}{ \partial e} 
  \\  
  &+ \lambda \beta_{\lambda} \frac{ \partial}{ \partial \lambda} 
  + \beta_{g} \frac{ \partial}{ \partial g} 
  +\sum_{n=1}^{3} \beta_{\xi_n} \frac{ \partial}{ \partial \xi_n} 
  + m^2 \delta \frac{ \partial}{ \partial m^2} 
  - \gamma_{\Gamma}
  \Bigg] 
  \Gamma^{(N)} 
  = 0,
\end{aligned}
  \label{RGEquations}
\end{eqnarray}
where $2 \gamma_{\Gamma} = N_{\phi} \gamma_{\phi}+N_{A_{0}} \gamma_{A_{0}}+N_{A_{i}}\gamma_{A_{i}}$ and
\begin{eqnarray}
  \begin{aligned}
   \beta_{a_1} &= \frac{\mu}{a_1} \frac{da_1}{d \mu} \, , \;   
   \beta_{a_2} =  \mu \frac{da_2}{d \mu} \; , \;
   \beta_{b_1} =  \frac{\mu}{b_1} \frac{db_1}{d \mu} \, , \;  
   \beta_{b_2} =  \mu \frac{db_2}{d \mu} \, , \; 
   \\ 
   \beta_{b_3}& =  \mu \frac{db_3}{d \mu} \, , \; 
   \beta_{b_4} =  \mu \frac{db_4}{d \mu} \, , \; 
   \beta_{e} =  \mu \frac{de}{d \mu} \, , \;       
   \beta_{\lambda} =  \frac{\mu}{\lambda} \frac{d\lambda}{d \mu} \, , \;  
      \beta_{g} =  \mu \frac{dg}{d \mu} \, , \;  
   \\  
      \beta_{\xi_n} &=  \mu \frac{d\xi_n}{d \mu} \, , \quad  
      \delta =  \frac{\mu}{m^2} \frac{dm^2}{d \mu}   \quad \text{and} \quad
   \gamma_{\Gamma} = \frac{ \mu}{Z_{\Gamma}} \frac{dZ_{\Gamma}}{d \mu}.
 \end{aligned}
   \label{RGFunctionsDef}
\end{eqnarray}

To obtain the above functions we proceed as follows. 
 We substitute the vertex  functions listed in the previous section in the renormalization group equation
and equate to zero the coefficient of each power of the momentum and each power of the coupling constants. In the case of the pure gauge functions, for instance, we determine

\begin{equation}
  \beta_{a_1} = \frac{\alpha}{16} \left[ \frac{R + a_1^2}{a_1^2} \right]
\qquad \text{and} \qquad
  \beta_{a_2} = \frac{\alpha}{16 } \left[ \frac{S + a_2^2}{a_2} \right],
  \label{Betaa1a2}
\end{equation}
and also

\begin{equation}
  \gamma_A \equiv \gamma_{A_{0}}=\gamma_{A_{i}}= \frac{\alpha}{8}.
  \label{GammaA}
\end{equation}
Furthermore, by inserting the scalar field two point function into the renormalization group equation we get
\begin{equation}
  \delta = \frac{1}{2} \left[ \frac{Q_3 - \alpha m^2}{m^2} \right],
  \label{Delta}
\end{equation}

\begin{equation}
  \gamma_{\phi} = - \, \frac{\alpha}{2}
  \label{GammaPhi}
\end{equation}
and

\begin{equation}
  \beta_{b_1} = \frac{1}{4} \left[ \frac{Q_1 -\alpha  b_1^2}{b_1^2} \right]
\qquad \text{and} \qquad
  \beta_{b_2} = \frac{1}{4} \left[ \frac{Q_2 -\alpha  b_2^2}{b_{2}^{2}} \right].
  \label{Betab1b2}
\end{equation}

Similarly, using the three point vertex function, we get
\begin{equation}
\beta_e= \frac{e\alpha}{16}, \qquad\qquad\beta_{b_{3}}= \frac{1}{4}\Big[\frac{ K_{3}-\alpha b_{3}^{2}}{b_{3}^{2}}\Big].
\end{equation}

Even without calculating the radiative corrections for the vertices with more than three fields,  the results obtained so far, together with some reasonable assumptions, allow  us to examine  relevant questions related to the possible  emergence  of Lorentz symmetry at  low energies.
For that purpose, we recall that, as a function of the momenta and the parameters of the model, $\Gamma^{(N)}$ has dimension $6- N_{\phi}-N_{A_{i}}-2N_{A_{0}}$ and therefore satisfies
\begin{equation}
\bigg[ 2 p_{0}\frac{\partial}{\partial\, p_{0}}+p\frac{\partial}{\partial\, p}+
  2\mu \frac{ \partial}{ \partial \mu}  
  +a_1  \frac{ \partial}{ \partial a_1} 
  +  b_1  \frac{ \partial}{ \partial b_1} 
  + 2\lambda \frac{ \partial}{ \partial \lambda} 
  +4 m^2  \frac{ \partial}{ \partial m^2} 
  - (6-N_{\phi}-N_{A_{i}}-2N_{A_{0}})
  \Bigg] 
  \Gamma^{(N)} 
  = 0,
  \label{DIMequations}
\end{equation}
where $p_{0}$  and $p $ symbolically stand for the sets of  time like  and space like parts of the momenta. 
 From (\ref{DIMequations}) and the renormalization group equation we may now write,
\begin{eqnarray}
\begin{aligned}
&\bigl [-\frac{\partial}{\partial t}+(\beta_{a_{1}}-\frac{1}{2})a_{1}\frac{\partial}{\partial a_1}+(\beta_{b_{1}}-\frac{1}{2})b_{1}\frac{\partial}{\partial b_{1}}+(\beta_{\lambda}-1)\lambda\frac{\partial}{\partial \lambda}+\beta_{a_{2}}\frac{\partial}{\partial a_{2}}\bigr.\\
&+\sum_{i=2}^{4}(\beta_{b_{i}}\frac{\partial}{\partial b_{i}}) +\sum_{n=1}^{3}\beta_{\xi_{n}}\frac{\partial}{\partial \xi_{n}}+\beta_{e}\frac{\partial}{\partial_{e}}+\beta_{g}\frac{\partial}{\partial_{g}}+(\delta-2)m^2\frac{\partial}{\partial m^{2}}\\
&\bigl.+\frac{1}{2}(6-N_{\phi}-N_{A_{i}}-2N_{A_{0}})-\gamma_{\Gamma}\bigr ]\Gamma^{(N)}(e^{t}p_{0},e^{t/2}p,x)=0,
\end{aligned} 
 \end{eqnarray} 
 where $x$ designates the set of  parameters of the model, specified in (\ref{Lagrangian}) and (\ref{Lagrangian1}).  To solve this equation, we introduce  running couplings. For the coefficients of the renormalizable (marginal) vertices, generically denoted by $\bar a(a,t)$,  they obey 
 \begin{equation}
 \frac{\partial\bar a}{\partial t}=\beta_{\bar a}
 \end{equation}
and the  initial condition $\bar a(a,0)=a$. On the other hand,  the running couplings associated with  the coefficients  of the super-renormalizable (relevant) vertices 
${\bar m}(m,t)$, $\bar a_{1}(a_{1},t)$, $\bar b_{1}(b_{1},t)$ and $\bar{\lambda}(\lambda,t)$ must satisfy
\begin{equation}
\frac{\partial {\bar m}^{2}}{\partial t}=(\delta -2){\bar m}^{2},\quad\frac{\partial \bar a_{1}}{\partial t}=(\beta_{\bar a_{1}}-\frac{1}{2})\bar a_{1},\quad\frac{\partial \bar b_{1}}{\partial t}=(\beta_{\bar b_{1}}-\frac{1}{2})\bar b_{1},\quad
\frac{\partial \bar \lambda}{\partial t}=(\beta_{\bar \lambda}-1)\bar \lambda, 
\end{equation}
also subject to the condition  that at $t=0$ they are equal to the original parameters.
 Thus, for the couplings $a_{2}$, $b_{i}$ with $i=2,3,4$ and $\xi_{n}$, with $n=1,2,3$, Lorentz symmetry demands that the corresponding $\beta$ functions be positive for small energies. This however will not be enough if
$a_{1}\not = b_{1}$. Thus, we set $a_{1}=b_{1}=c$ as a starting condition for these parameters in the original Lagrangian and require $\beta_{\bar a_{1}}=\beta_{\bar b_{1}}$ so that they remain equal as $t $ varies. 

We may now factorize $c^{2}$ out from the Lagrangian and redefine $c^{-1}\partial_{0}\rightarrow\partial_{0}$,  $c^{-1}A_{0}\rightarrow A_{0}$, $c^{-2} m^2 \rightarrow m^2$, $c^{-2} \lambda \rightarrow \lambda$ e $c^{-2} g \rightarrow g$. We get a new Lagrangian with the usual
terms of the $4+1 $ scalar QED and	with the high derivative terms divided by $c^{2}$. For the emergence of the Lorentz symmetry to take place, the coefficients of these terms should be small. Let $a^{2}/c^{2}$ be one of these coefficients;   we shall have then
\begin{equation}
    \frac{\partial}{\partial t}\left(\frac{\bar a^2}{\bar c^2}\right)= \frac{2 \bar a}{\bar c^2}\left[\beta_{\bar a}-\bar a(\beta_{\bar c}-\frac{1}{2})\right]>0.  
    \label{BetaCombinations}
\end{equation}

One simplification is to set the $\xi_{n}=0$, assuming that at least to one loop they are not generated by the radiative corrections. The choice $a_{1}=b_{1}=c$ corresponds to the assumption that, in the absence of high derivatives terms, the speed of light  is well defined. The imposition that   $\beta_{\bar a_{1}}=\beta_{\bar b_{1}}$ implies that this velocity remains well defined although it may change with the energy. 
However, the system  of equations is still very complicated so we restrict our analysis  to the situation  in which $\beta_{\bar c}=0$. This condition
allows one to fix $b_{3}$ and $b_{4}$ as functions of $a_{2}$ and $b_{2}$:

\begin{equation}
  \begin{aligned}
  b_3 &= \sqrt{\frac{b_2}{3(a_2^2+a_2 b_2+b_2^2)}} 
  \bigg[ 3b_2 (3 a_2^2 - a_2 b_2 - b_2^2) 
  \\ & \quad 
  \pm \sqrt{3} \sqrt{-2a_2^6-4a_2^5 b_2 +8a_2^4 b_2 +7a_2^3 b_2^3 +18a_2^2b_2^4+18a_2b_2^5+9b_2^6} 
  \bigg]^{1/2}
\end{aligned}
  \label{b3Caso1}
\end{equation}
and
\begin{equation}
  \begin{aligned}
  b_4 &= \sqrt{\frac{b_2}{6(a_2^2+a_2 b_2+b_2^2)}} 
  \bigg[ 3b_2 ( a_2^2 - 7 a_2 b_2 - 7 b_2^2) 
  \\ & \quad 
  \pm 4 \sqrt{3} \sqrt{-2a_2^6-4a_2^5 b_2 +8a_2^4 b_2 +7a_2^3 b_2^3 +18a_2^2b_2^4+18a_2b_2^5+9b_2^6} 
  \bigg]^{1/2},
\end{aligned}
  \label{b4Caso1}
\end{equation}
with the use of the signs $+$ or $-$  in  these expressions to be discussed shortly.
By using (\ref{b3Caso1}), we can eliminate the dependence on $b_{3}$ and $b_{4}$ from  $\beta_{a_{2}}$, $\beta_{b_{2}}$ and $\beta_{b_{3}}$ so that they become
\begin{equation}
  \begin{aligned}
    \beta_{a2} &= \frac{e^2}{27648 \pi^2 a_2 b_2} \Bigg\{
    108 a_2^2 +27 b_2^2 + \frac{42 b_2 \left( 3b_2 P_1(a_2,b_2) \mp 2 \sqrt{3} \sqrt{P_2(a_2,b_2)} \right)}{P_3(a_2,b_2)} 
    \\ & \qquad
    + \frac{\left( 3b_2 P_1(a_2,b_2) \mp 2 \sqrt{3} \sqrt{P_2(a_2,b_2)} \right)^2}{\left(\, P_3(a_2,b_2) \, \right)^2 } 
    \Bigg\} ,
  \end{aligned}
  \label{Betaa2Caso1}
\end{equation}

\begin{equation}
  \begin{aligned}
    \beta_{b_2} &= \frac{e^2}{2304 \pi^2} \Bigg\{
    -36
    + \frac{1}{a_2 (a_2+b_2)^3} \Bigg[
    9 b_2^2 (23a_2^2+37a_2b_2 +16 b_2^2) 
    \\ & \qquad
    - \frac{a_2 (a_2 +3b_2)\left( 3 b_2 P_1(a_2,b_2) \mp 2 \sqrt{3} \sqrt{P_2(a_2,b_2)} \right)^2}{\left( \, P_3(a_2,b_2) \, \right)^{2}}
    \\ & \qquad \qquad
    - \frac{6b_2 (7 a_2^2 +9a_2 b_2 +4b_2^2)\left( 3 b_2 P_1(a_2,b_2) \mp 2 \sqrt{3} \sqrt{P_2(a_2,b_2) } \right)}{P_3(a_2,b_2)} 
    \Bigg]
    \Bigg\}
  \end{aligned}
  \label{Betab2Caso1}
\end{equation}
and

\begin{equation}
  \begin{aligned}
    \beta_{b_3} &= 
    \frac{e^2 P_3(a_2,b_2)}{2304 \pi^2 a_2 b_2(a_2+b_2)^3 
      \left( - 3b_2 P_1(a_2,b_2) \pm 2 \sqrt{3} \sqrt{P_2(a_2,b_2)} \right) }
    \\ &
    \times \Bigg\{
9 b_2 (3a_2^4 +9a_2^3 b_2 +25a_2^2b_2^2 +41a_2 b_2^3+20 b_2^4)
    \\ &
    + \frac{3(21a_2^4+ 63a_2^3 b_2+23a_2^2 b_2^2 -63 a_2 b_2^3 - 40b_2^4 )
    \left( 3b_2 P_1(a_2,b_2) \mp 2 \sqrt{3}\sqrt{P_2(a_2,b_2)} \right)}{P_3(a_2,b_2)}
    \\ & \qquad
    + \frac{2 b_2(8a_2^2+ 15a_2 b_2+6 b_2^2)
    \left( 3b_2 P_1(a_2,b_2) \mp 2 \sqrt{3}\sqrt{P_2(a_2,b_2)} \right)^2}{\left( \,P_3(a_2,b_2)
    \, \right)^2}
    \Bigg\},
  \end{aligned}
    \label{BetaFunctionsCaso1}
\end{equation}
where the polynomials $P_{1}(a_{2},b_{2})$, $P_{2}(a_{2},b_{2})$ and $P_{3}(a_{2},b_{2})$ were introduced to simplify the writing of the above expressions; they are given by

\begin{equation}
  \begin{aligned}
   P_1(a_2,b_2) & = -3 a_2^2+a_2 b_2+b_2^2,
   \\
   P_2(a_2,b_2) & = -2 a_2^6 - 4 a_2^5 b_2 + 8 a_2^4 b_2^2 + 7 a_2^3 b_2^3 + 18 a_2^2 b_2^4 
   + 18 a_2 b_2^5 + 9 b_2^6,
   \\
   P_3(a_2,b_2) & = a_2^2 + a_2 b_2 + b_2^2.
  \end{aligned}
    \label{Polinomios}
\end{equation}

Because of the complexity of these expressions we will employ numerical methods to find regions where the parameters decrease by lowering
the energy: firstly, we find  zeros of (\ref{BetaFunctionsCaso1}) and then analyze the
behavior of these functions as perturbed around the zeros. Then we do the same for
(\ref{Betaa2Caso1}) and (\ref{Betab2Caso1}) and obtain:

\begin{itemize}
  \item In the interval $0 \leq \frac{b_2}{a_2} < 0.62429879$, we have $\beta_{a_2} > 0$.
  \item In the interval $\frac{b_2}{a_2} > 0.48792827$, we have $\beta_{b_2} > 0$.
  \item In the interval $\frac{b_2}{a_2} > 0.49508332$, we have $\beta_{b_3} > 0$.
\end{itemize}

Finally, concerning the behavior of $\beta_{b_{4}}$ we notice that, as $b_{4}$ is a function of $a_{2}$ and $b_{2}$,
\begin{equation}
\beta_{b_4}
  = \beta_{a_2} \frac{\partial b_4}{\partial a_2}
  + \beta_{b_2} \frac{\partial b_4}{\partial b_2}.
  \label{Betab4Caso1}
\end{equation}
and performing the same analysis described above, we find that $\beta_{b_4}$ is positive
for $\frac{b_2}{a_2} > 0.50848002$. Thus, by collecting all these results, we find that in the  interval 

\begin{equation}
  0.50848002 < \frac{b_2}{a_2} < 0.62429879
  \label{LorentzRestoringCaso1Final}
\end{equation}
all $\beta$ functions are positive.  Lorentz symmetry may emerge but this requires a fine tuning procedure, as described.

\section{Summary and conclusions}\label{section5}
In this work we studied the $z=2$ scalar quantum electrodynamics in five spacetime dimensions. We regularized the Feynman amplitudes by promoting the spatial part of the Feynman integrals to $d=4-\epsilon$ and the renormalization of the model was accomplished by removing the pole
parts of the  result (minimum subtraction procedure). We explicitly checked that these pole parts satisfy, as they should,  the  Ward  identities characteristics of the model. By determining the  relevant $\beta$ functions,  we analyzed  possible scenarios for the evolution of  various coupling constants.  We verified  that the emergence of the Lorentz symmetry may occur in the low energy limit but this requires a fine tuning procedure. Another possibility is to have 
an ultraviolet regime in which the usual, quadratic terms  in the derivatives, become negligible. This is a great simplification making possible
to determine completely the one-loop integrals. As a  third scenario, there is the opposite situation where the usual terms may be very large, which would be interesting for applications to the physics of the early universe.

Finite temperature/density effects may be considered using standard methods. In particular,
that extension does not significantly alter the ultraviolet structure we analyzed in this work.
This is so because, the conserved charge density  still has the usual form,
 \begin{equation}
 j_{0}(x)= i(\phi^{*} D_{0}\phi-\phi(D_{0}\phi)^{*})
 \end{equation}
 and therefore  the chemical potential vertex $\mu_{0}j_{0}$ is super-renormalizable.  Its  impact on the UV behavior manifests itself through the inclusion of  a term $-2 \nu_{\mu_{0}}$ in the power counting. A  finite temperature $T$ may also be considered by discretizing the temporal part of the momenta through the Matsubara replacement $p_{0}\rightarrow  (2 i\pi n) T$. Of course, the resulting ultraviolet structure is the same as before the replacement. 

Similarly,  to make contact with the four dimensional physics, we may compactify one spatial dimension in a circle. Imposing to the fields periodic boundary conditions in that fifth dimension, we get towers of Kaluza-Klein modes of increasing masses. As in the case of finite temperature, this construction does not alter the ultraviolet structure discussed in this work. The phenomenological aspects of this structure,  have not been treated here and will be  the subject of future work.

\section{Acknowledgments}

We would like to thank Dr. E. A. Gallegos for helpful discussions in the early stages of this research.
This work was partially supported by Conselho Nacional de Desenvolvimento  Cient\'{\i}fico e Tecnol\'ogico (CNPq).
\appendix
\section{One-loop corrections and renormalization}\label{appendixA}
In this Appendix we will examine the ultraviolet structure of the model by analyzing the possible divergences as specified by (\ref{1}). As mentioned before, our Feynman integrals are dimensionally regulated by promoting their spatial parts to $d=4-\epsilon$ dimensions. These integrals are not analytically feaseble and to extract their divergent parts we  Taylor expand their integrands in powers of the external momenta. For a Feynman amplitude $I_{\Gamma}$
of a graph $\Gamma$ we use
\begin{equation}
I_{\Gamma}(p) = \sum_{s=0}^{[\frac{\delta(\Gamma)}{2}]}\frac{p_{0}^{s}}{s!}\frac{\partial^{s}{\phantom a}}{\partial p_{0}^{s}}\sum_{n=0}^{\delta(\Gamma)-2s}\frac{{ p}_{i_{1}}\ldots{ p}_{i_{n}}}{n!}\frac{\partial\phantom a}{\partial{ p}_{i_{1}}}\ldots\frac{\partial\phantom a}{\partial{ p}_{i_{n}}}I_{\Gamma} + \text{finite part},\label{2}
\end{equation}
where $\delta(\Gamma)$ is the degree of superficial divergence of $\Gamma$, $[x]$ is the greatest integer less than or equal to $ x $, $ p_{0}^{s} $ symbolically stands for the product of $ s $ timelike components of an independent set of external momenta; ${ p}_{i}$ denotes the ith spacelike component and all derivatives are computed at zero external momenta. 

By using (\ref{2}), we obtain for the coefficients of the Taylor expansion, integrals of the type

\begin{equation}
  J(x,y,z) = \int \frac{dk_{0}}{2\pi}\frac{d^dk}{(2\pi)^d} 
  \frac{k_0^x |\vec{k}|^y}{[k_0^2 -b_1^2 \vec{k}^2 -b_2^2 \vec{k}^4 - m^2]^z}
  \label{IntegralPedro}
\end{equation}
or
\begin{equation}
\int \frac{dk_0}{2\pi}\frac{d^dk}{(2\pi)^d}\frac{1}{(k_{0}^{2}-b^{2}_{1} {\vec k}^2-a^{2}_{1} ({\vec k}^2)^2-m^{2}_{1})^{z_1}} \frac{1}{(k_{0}^{2}-b^{2}_{2} {\vec k}^2-a^{2}_{2} ({\vec k}^2)^2-m^{2}_{2})^{z_2}},
\end{equation}
if there are propagators with different denominators in the loop integral. In this last case,  we use Feynman's trick
\begin{equation}
\frac{1}{A^{z_{1}}B^{z_{2}}}= \frac{\Gamma(z_{1}+z_{2})}{\Gamma(z_{1})\Gamma(z_{2})}\int_{0}^{1} dx\, \frac{x^{z_{1}-1}(1-x)^{z_{2}-1}}{[A x+ B(1-x)]^{z_{1}+z_{2}}}
\end{equation}
to obtain an integral similar to (\ref{IntegralPedro}).
 The divergent part of this integral may be calculated using standard methods (see appendix in \cite{Gomes1} for details) yielding the result
 
 \begin{equation}
  \begin{aligned}
    J(x,y,z) = \frac{i^{1+x-2z}}{(4\pi)^{(d+2)/2}}  &
  \frac{\left[(-1)^x+1\right]}{2}
  \frac{ \Gamma \left( \tfrac{x+1}{2} \right)}{ \Gamma \left( \tfrac{d}{2} \right) \Gamma (z)}
  \\  & \quad \times
  \sum_{n=0}^{2} \frac{(- b_1^2)^n}{n!}
  \frac{ \Gamma \left( \tfrac{d+y+2n}{4} \right) \Gamma \left( \omega + \tfrac{(n-1)}{2} \right)}{(b_2^2)^{(d+y+2n)/4}(m^2)^{\omega + \frac{(n-1)}{2}}},
\end{aligned}
  \label{FormulaFinalIntegralPedroAppendixSimplificada}
\end{equation}
where $w=(4z-2x-y-d)/4$.
We now analyze the possible divergences  on the effective action, as indicated  in  (\ref{1}).
We have

1. Pure gauge sector, i.e., graphs with $N_{\phi}=0$. In what follows, $\Pi_{\mu\nu}$ will denote the correction to the  kernel of the term with two gauge fields in the effective action, i.e., the term $A^{\mu}(p) \Pi_{\mu\nu}A^{\nu}(-p)$. Due to gauge invariance, the counterterms must depend on the potential $A_{\mu}$ only through the gauge field strength $F_{\mu\nu}$. Also, charge conjugation symmetry restricts the number of external lines to be even.

 For $N_{A_{0}}=2$  the divergences are quadratic. Using the Feynman rules stated before, we found the following contributions coming from the graphs depicted in Fig 1

\begin{equation}
    \Pi_{00}(p) =e^2 \Bigg [-   
    \int [dk] \frac{1}{\Omega_b[k^2]} +\frac{1}{2}  
    \int [dk] 
  \frac{(p_0 + 2 k_0 )^2}{\Omega_b[k^2]\Omega_b[(k+p)^2]}\Bigg ],
  \label{Pi003}
\end{equation}
where here and henceforth we employ the notation $\Omega_{b}(k)\equiv k_{0}^{2}-b_{1}^{2}{\vec k}^{2}-b_{2}^{2}{\vec k}^{4}-m^{2}$ and $[dk]\equiv \mu^{\epsilon/2} dk_{0}d^{d}k/(2\pi)^{d+1}$. Due to the presence of quadratic and quartic terms in the denominators of the  integrands, the above integrals does not produce simple analytic expressions. The pole part of the result may be nevertheless easily computed by expanding the  integrands in power series and using (\ref{FormulaFinalIntegralPedroAppendixSimplificada}) as described before. Proceeding in this way we found
\begin{equation}
  \Pi_{00}(p) = \frac{i}{4}\alpha  \mu^{\epsilon/2}  \Bigg[ 
    \frac{1}{ \epsilon}\vec{p}^{\,2} 
     + \text{(finite part)} \Bigg],
  \label{GaugeCorrectionResidue00}
\end{equation} 
where $\alpha=\frac{e^2}{16 \pi^2 b_2}$.
For $N_{A_{0}}=1$ and $N_{A_{i}}=1$ the graphs have same topology as before but different polynomials at the vertices. We have
\begin{eqnarray}
 \begin{aligned}
    \Pi_{0i}(p)&= -\frac{e^2}{4}  
    \Bigg\{
  b_1^2 \int [dk] 
  \frac{(p_0 + 2 k_0 )(p_i + 2 k_i)}{\Omega_b[k^2]\Omega_b[(k+p)^2]} \\
  &
  + b_2^2 \int [dk] 
  \frac{(p_0 + 2 k_0 )(p_l + 2 k_l)\{ (p_i + k_i )(p_l + k_l ) + k_i k_l \}}{\Omega_b[k^2]\Omega_b[(k+p)^2]} 
  \Bigg\},
\end{aligned}
  \label{Pi0i3}
\end{eqnarray}
yielding
\begin{equation}
  \Pi_{0i}(p) = \Pi_{i0}(p) = \frac{i}{4}\alpha  \mu^{\epsilon/2} \Bigg[ 
    \frac{1}{ \epsilon}p_0 p_i
     + \text{(finite part)} \Bigg].
  \label{GaugeCorrectionResidue0i}
\end{equation}

c. Similarly, for 
$N_{A_{i}}=2$ divergences arise only if $\nu_{3B}=0$ and in that case the degree of superficial divergence is four.  Explicit calculation gives

\begin{eqnarray}
    \Pi_{ij}(p) &=&- e^2  \delta_{ij} \Bigg\{ 
  (b_1^2 + b_2^2 \vec{p}^{\,2} ) 
  \int [dk] \frac{1}{\Omega_b[k^2]} 
  + \frac{2}{d(d+2)} b_2^2 
  \int [dk] \frac{\vec{k}^{\,2}}{\Omega_b[k^2]} \Bigg\}\nonumber\\
&&+     
    \frac{e^2}{2} 
 \int [dk]  \Bigg\{
  b_1^4  \frac{(p_i + 2 k_i )(p_j + 2 k_j)}{\Omega_b[k^2]\Omega_b[(k+p)^2]}\nonumber \\
 && +  b_2^4  
  \frac{(p_l + 2 k_l )(p_m + 2 k_m)\{ (p_i + k_i )(p_l + k_l ) + k_i k_l \} \{ (p_j + k_j )(p_m + k_m ) + k_j k_m \}}{\Omega_b[k^2]\Omega_b[(k+p)^2]} \nonumber\\
 && +  2 b_1^2 b_2^2 
  \frac{(p_i + 2 k_i )(p_l + 2 k_l)\{ (p_j + k_j )(p_l + k_l ) + k_j k_l \}}{\Omega_b[k^2]\Omega_b[(k+p)^2]} 
  \Bigg\},
  \label{Piij3}
\end{eqnarray}

so that
\begin{eqnarray}
  \begin{aligned}
    \Pi_{ij}(p) =& \frac{i\alpha\mu^{\epsilon/2}}{4}   
    \Bigg[ \,
    \frac{1}{ \epsilon} 
    \bigg(
    \delta_{ij} p_0^2 
    + R (\delta_{ij} \vec{p}^{\,2} - p_i p_j)
    +S (\delta_{ij} \vec{p}^{\,2} - p_i p_j) \vec{p}^{\,2}
  \bigg) \\
     & + \text{(finite part)} \Bigg],
  \end{aligned}
\end{eqnarray}
where 
\begin{equation}
  R = \frac{2 b_1^2}{b_2^2}(2 b_2^2 - b_3^2 +b_4^2)
  \qquad \text{and} \qquad
  S = \frac{1}{12 b_2^2} (3 b_2^4 - 14 b_2^2 b_3^2 + b_3^4) .
  \label{RS}
\end{equation}
Thus the counterterms have the forms $C_{1}F_{i0}F_{i0}$, $C_{2}F_{ij}F_{ij}$ and $C_{3}\partial_l F_{i j} \partial_l F_{i j} $, where
\begin{equation}
C_{1}=\frac{\alpha}{8}\frac{1}{\epsilon},\quad
C_{2}= \frac{\alpha}{4}\frac{b_{1}^{2}}{b_{2}^{2}}\frac{(2b_{2}^{2}-b_{3}^{2}+b_{4}^{2})}{\epsilon}, \quad C_{3}=\frac{\alpha}{96 b_{2}^{2}}\frac{(3 b_{2}^{4}-14b_{2}^{2}b_{3}^{2}+b_{3}^{4})}{\epsilon}.
\end{equation}
 The  above results also show that the wave renormalization functions for the fields $A_{0}$ and $A_{i}$ are equal. Observe  that  for $b_{1}=0$ there is no contribution to the term $F_{ij}F_{ij}$, as expected because of  conformal invariance.

d. $N_{A_{i}}=4$. Here the relevant graphs are quadratically divergent but, as the counterterms  necessarily
 depend on the potential only through the field strength, four momentum factors are needed to produce a nonzero result. In this case, the contribution is finite.
   
2. Matter/gauge field mixed sector. Firstly we have $N_{A}=0$  and $N_{\phi}=2$. In this case we found one-loop corrections of the form $\phi^{*}\Delta \Gamma^{(2)}\phi$, coming from graphs with three different topologies, as shown in Fig 2. The tadpole graphs, Fig 2a, have one  internal scalar line and the vertex is either the $\lambda$ vertex or one of the vertices with the couplings $\xi_{n}$.    They furnish
\begin{equation}
 \Delta \Gamma_{1}^{(2)}=\lambda \int [dk]  
  \frac{1}{\Omega_b[k^2]}+ (4 \xi_1 - \xi_3) \left\{ 
  \int [dk] \frac{\vec{k}^2}{\Omega_b[k^2]} 
  + \vec{p}^2 \int [dk] \frac{1}{\Omega_b[k^2]} 
  \right\} .
\end{equation}
There is a also a tadpole graph with internal spatial gauge field propagator as shown in Fig 2b (notice that, since the spatial part is dimensionally regularized, the would be contribution of the tadpole graph with internal time like gauge propagator vanishes),
\begin{equation}
\Delta \Gamma_{2}^{(2)}=-e^{2}(1-d)\int[dk]\frac{b_{1}^{2}+b_{2}^{2}(k^{2}+\frac{4}{d}{\vec p}^2)}{\Omega_{a}[k^2]}.
\end{equation}
There are,  finally, the  contributions from the graphs with two trilinear vertices, see Fig. 2c.
\begin{equation} 
\begin{aligned}
\Delta \Gamma_{3}^{(2)}=&e^2\int [dk] \Bigg\{
   \frac{(k_0 + 2 p_0 )^2}{\vec{k}^2 \Omega_b[(k+p)^2]}
   +   
   \frac{4 b_1^4 p_i p_j \left( \delta_{ij}- \tfrac{k_i k_j}{ \vec{k}^2} \right)}{\Omega_a[k^2]\Omega_b[(k+p)^2]} \\
  + &
  \frac{ b_2^4 p_i p_j (\vec{k} + 2 \vec{p} )^4 \left( \delta_{ij}- \tfrac{k_i k_j}{ \vec{k}^2} \right)}{\Omega_a[k^2]\Omega_b[(k+p)^2]} 
  +  
  \frac{4 b_1^2 b_2^2 p_i p_j (\vec{k} + 2 \vec{p} )^2 \left( \delta_{ij}- \tfrac{k_i k_j}{ \vec{k}^2} \right)}{\Omega_a[k^2]\Omega_b[(k+p)^2]}
  \Bigg\}.
\end{aligned}
  \label{Xi3}
\end{equation}
By performing the indicated integrals in the above expressions, we obtain the total correction, $\Delta \Gamma^{(2)}=\Delta \Gamma_{1}^{(2)}+\Delta \Gamma_{2}^{(2)}+\Delta \Gamma_{3}^{(2)}$, to the two point vertex function for the scalar field,

\begin{equation}
 \Delta   \Gamma^{(2)} =  i  \mu^{\epsilon/2}
 \Bigg[ 
    \frac{1}{ \epsilon} 
    \bigg(
    -\alpha p_0^2 + Q_1 \vec{p}^{\, 2} + Q_2 \vec{p}^{\,4} + Q_3  
    \bigg)
    + \text{(finite part)} \Bigg],
  \label{MatterCorrectionResidue}
\end{equation}
where $Q_{i}$ with $i=1,2,3$ are
\begin{eqnarray}
    Q_1 &=&\frac{3\alpha}{8  a_2^3 b_2^2 (a_2 + b_2)^2}  
    \bigg\{
    2 a_1^2 a_2 b_2^2 (11 b_2^4 -2 b_2^2 b_3^2 - b_3^4 )
    + a_1^2 b_2^3 (11 b_2^4-2 b_2^2 b_3^2 - b_3^4 ) \nonumber
    \\ && 
    + a_2^3 b_1^2 (7 b_2^4 + 6 b_2^2 b_3^2-b_3^4 )
    +2 a_2^2 b_2 (6 a_1^2 b_2^4+b_1^2 (3 b_2^4+2 b_2^2 b_3^2-b_3^4))
    \bigg\}\nonumber\\
   && + \left( 4 \xi_1 - \xi_3 \right) \frac{b_1^2}{32 \pi^2 b_2^3} ,
    \label{Q1}
\end{eqnarray}
\begin{eqnarray}
  \begin{aligned}
    Q_2 =&  \frac{\alpha}{4 a_2  (a_2 + b_2)^3} 
    \bigg\{
    b_2^4 (23 a_2^2+37 a_2 b_2+16 b_2^2)
    +2 b_2^2 (7 a_2^2+9 a_2 b_2+4 b_2^2) b_3^2 \\
    & 
    -a_2 (a_2+3 b_2) b_3^4
    \bigg\}
    \label{Q2}
  \end{aligned}
\end{eqnarray}
and
\begin{eqnarray}
  \begin{aligned}
    Q_3 =&  \frac{\alpha}{8  a_2^5 b_2^2 } 
    \bigg\{
    12 a_1^2 a_2^2 b_1^2 b_2^3
    -9 a_1^4 b_2^3 (b_2^2+b_4^2)
    +a_2^5 b_1^4 
    - 4 a_2^5 b_2^2 m^2
    \bigg\}
    \\ &  
    + \lambda \frac{b_1^2}{32 \pi^2 b_2^3} 
    + \left( 4\xi_1 - \xi_3 \right) \frac{ \left( 4b_2^2m^2 -3b_1^2 \right) }{32 \pi^2 b_2^5}.
    \label{Q3}
  \end{aligned}
\end{eqnarray}
It should be noted that, for $a_{1}=b_{1}=0$,  $Q_{1}$ vanishes so that corrections to the lowest order terms in the spatial derivatives do not occur. Observe  that, if also $m=0$, then $Q_{3}$ vanishes so that conformal invariance is preserved.

3. Three point vertex function associated with the product $\phi(p)\phi^{*}(-p')A_{\mu}(p-p')$(see graphs in Fig 3). 

3.a The contributions for the correction for the vertex $V_{3A}$ ($N_{A_{0}}=1$ and $N_{\phi}=2$ ) was found to be 
\begin{equation}
\Delta\Gamma^{(3)0}_{1}= e^3\int [dk]\left[\frac{p_{0}}{(\vec{k}+\vec{p})^2\Omega_{b}[k^2]}+\frac{p'_{0}}{(\vec{k}+\vec{p'})^2\Omega_{b}[k^2]}\right],
\end{equation}
coming from the graphs with two vertices and
\begin{equation}
  \begin{aligned}
   \Delta\Gamma^{(3)0}_{2}= & i e^3 
    \int [dk] (p_0 + p'_0 + 2 k_0) 
    \Bigg\{ 
  \mbox{\Large$ \frac{(k_0 + 2 p_0 )(k_0 + 2 p'_0)}{\vec{k}^2 \Omega_b[(k+p)^2] \Omega_b[(k+p')^2]}$}  
   +  
   \mbox{\Large$ \frac{ 4 b_1^4 p_i p'_j \left( \delta_{ij}- \tfrac{k_i k_j}{ \vec{k}^2} \right)}{\Omega_a[k^2]\Omega_b[(k+p)^2]\Omega_b[(k+p')^2]}$} \\ 
  & \qquad \qquad +
  \mbox{\Large$ \frac{ b_2^4  p_i p'_j (\vec{k} + 2 \vec{p} )^2 (\vec{k} + 2 \vec{p}\,')^2 \left( \delta_{ij}- \tfrac{k_i k_j}{ \vec{k}^2} \right)}{\Omega_a[k^2]\Omega_b[(k+p)^2]\Omega_b[(k+p')^2]}$} \\
  + &
  \mbox{\Large$ \frac{ 2 b_1^2 b_2^2 p_i p'_j (\vec{k} + 2 \vec{p}\,' )^2 \left( \delta_{ij}- \tfrac{k_i k_j}{ \vec{k}^2} \right)}{\Omega_a[k^2]\Omega_b[(k+p)^2]\Omega_b[(k+p')^2]}$} 
  + 
  \mbox{\Large$ \frac{2 b_1^2 b_2^2 p_i p'_j (\vec{k} + 2 \vec{p} )^2 \left( \delta_{ij}- \tfrac{k_i k_j}{ \vec{k}^2} \right)}{\Omega_a[k^2]\Omega_b[(k+p)^2]\Omega_b[(k+p')^2]}$}
  \Bigg\},
\end{aligned}
  \label{Lambda0333}
\end{equation}
coming from graphs with three vertices.  After performing the integrations, we obtain 
\begin{equation}
   \Delta\Gamma^{(3)0}(p,p') \equiv \Delta\Gamma^{(3)0}_{1}(p,p')+\Delta\Gamma^{(3)0}_{2}(p,p')=  i\frac{e^3\mu^{\epsilon/2}}{16 \pi^2 b_2} 
    \Bigg[ 
    - \frac{1}{ \epsilon} 
   \big(p^0+{p'}^{0} \big) 
    + \text{(finite part)} \Bigg].
  \label{3PointsGreenCorrectionResidueLambda0}
\end{equation}

3.b The divergent contribution to the three point vertex function with  spatial $A_{i}$ is more cumbersome. It  involves  graphs (ten with three vertices, six with two vertices and four tadpoles). The final result is
\begin{eqnarray}
\Delta\Gamma^{(3)i}   (p,p')& =&  ie\mu^{\epsilon/2}
     \Bigg\{
    \frac{1}{ \epsilon}
    \Big[ 
    K_1 \big( \, p_i + p'_i \, \big) \nonumber\\
    & &+K_2 \big( \, p_i (\vec{p}^{\,2} + \vec{p} \cdot \vec{p'} ) 
    + p'_i ( \vec{p'}^{\,2} + \vec{p} \cdot \vec{p'} ) \, \big)
     \nonumber\\
   & &+K_3 \big (p_i ( \vec{p'}^{\,2} - \vec{p} \cdot \vec{p'}  ) 
    + p'_i ( \vec{p}^{\,2} - \vec{p} \cdot \vec{p'}  ) \, \big) 
   \Big]
   + \text{(finite part)} \Bigg\},
  \label{3PointsGreenCorrectionResidueLambdai}
\end{eqnarray}
where
\begin{eqnarray}
 \begin{aligned}
    K_1 &=\frac{3 \alpha }{8 a_2^3 b_2^2 (a_2 + b_2)^2}  
    \bigg\{
    2 a_1^2 a_2 b_2^2 (11 b_2^4 -2 b_2^2 b_3^2 - b_3^4 )
    + a_1^2 b_2^3 (11 b_2^4-2 b_2^2 b_3^2 - b_3^4 ) 
    \\ & 
    + a_2^3 b_1^2 (7 b_2^4 + 6 b_2^2 b_3^2-b_3^4 )
    +2 a_2^2 b_2 (6 a_1^2 b_2^4+b_1^2 (3 b_2^4+2 b_2^2 b_3^2-b_3^4))
    \bigg\}
    \\ & 
    + \frac{\left( 4 \xi_1 - \xi_3 \right)b_{1}^{2} }{32\pi^2 b_2^3} ,
    \label{K1}
  \end{aligned}
\end{eqnarray}

\begin{eqnarray}
  \begin{aligned}
  K_2 &= \frac{ \alpha}{4 a_2 (a_2 + b_2)^3} 
    \bigg\{
    b_2^4 (23 a_2^2+37 a_2 b_2+16 b_2^2)
    +2 b_2^2 (7 a_2^2+9 a_2 b_2+4 b_2^2) b_3^2 \\
    & 
    -a_2 (a_2+3 b_2) b_3^4
    \bigg\} 
    \label{K2}
  \end{aligned}
\end{eqnarray}

and

\begin{eqnarray}
 \begin{aligned}
    K_3 &= 
    \frac{\alpha}{12 a_2 (a_2 + b_2)^3}
    \bigg\{
    3 a_2^4 b_2^2+9 a_2^3 b_2^3+25 a_2^2 b_2^4+41 a_2 b_2^5+20 b_2^6-3 a_2^4 b_3^2
    -9 a_2^3 b_2 b_3^2 
    \\  &  
    +31 a_2^2 b_2^2 b_3^2 +81 a_2 b_2^3 b_3^2+40 b_2^4 b_3^2+16 a_2^2 b_3^4
    +30 a_2 b_2 b_3^4+12 b_2^2 b_3^4
    \bigg\} 
    \\  &  
    + \frac{(3b_3^2-7b_2^2) \left( 4 \xi_2 - \xi_3 \right) }{192 \pi^2 b_2^3 }.
    \label{K3}
  \end{aligned}
\end{eqnarray}

These results allow us to prove the simplest Ward identities of the model, namely,
 
 \begin{equation}
  p_{\mu} \Pi^{\mu\nu} = 0  
  \qquad \text{and} \qquad
  ( p'_{\mu} - p_{\mu} ) \Gamma^{(3)\,\mu} = e \left[ \Gamma^{(2)}(p') - \Gamma^{(2)}(p) \right],
  \label{WardIdentities}
\end{equation}
where $\Gamma^{(2)}(p)$ is the two point vertex function of the scalar fields and 
$\Gamma^{(3)\,\mu} $   denotes the three point vertex function of the product of fields $A^\mu(p'-p)\phi^{*}(-p')\phi(p)$.
The first identity may be verified straightforwardly using the previous results for the components of the polarization tensor. 
It shows that the radiative correction to the gauge field two point function is transversal; in the tree approximation that function has also a longitudinal part due to the gauge fixing.  The second identity may also be verified using that, before renormalization,

\begin{equation}
\begin{aligned}
\Gamma^{(2)}=& i \, [ p_{0}^{2}-b_{1}^{2}{\vec{p}}^{2}-b_{2}^{2}{\vec{p}}^{4}- m^2 ]+ \Delta \Gamma^{(2)}\\
=&i\left[\left( 1 - \frac{\alpha}{ \epsilon} \right) p_0^2
    - \left( b_1^2 - \frac{1}{ \epsilon} Q_1 \right) \vec{p}^{\,2}
    - \left( b_2^2 - \frac{1}{ \epsilon} Q_2  \right) \vec{p}^{\,4}\right . \\
   - &\left . \left( m^2 - \frac{1}{\epsilon} Q_3 \right) 
    + (\text{finite part}) \right]
    \end{aligned}
\end{equation}
and 
\begin{equation}
  \begin{aligned}
    \Gamma^{(3)\,0} = i e[\left [\left( 1 - \frac{\alpha}{ \epsilon} \right) (p_0 + p'_0)+
   (\text{finite part})\right ],
  \end{aligned}
  \label{Final1PI3Points0}
\end{equation}
\begin{equation}
  \begin{aligned}
     \Gamma^{(3)\,i} &=    
    -ie \left[ \left( b_1^2 - \frac{1}{ \epsilon} K_1 \right) (p_i + p'_i)
    +  \left( b_2^2 - \frac{1}{ \epsilon} K_2  \right) 
    \big( \, p_i (\vec{p}^{\,2} + \vec{p} \cdot \vec{p'} ) 
    + p'_i ( \vec{p'}^{\,2} + \vec{p} \cdot \vec{p'} ) \, \big) \right.\\
    & \left.
    + \left( b_3^2 - \frac{1}{ \epsilon} K_3 \right)
    \big( \, p_i ( \vec{p'}^{\,2} - \vec{p} \cdot \vec{p'}  ) 
    + p'_i ( \vec{p}^{\,2} - \vec{p} \cdot \vec{p'}  ) \, \big) +
    (\text{finite part})\right].
  \end{aligned}
  \label{Final1PI3Pointsi}
\end{equation}
where, the expressions for $Q_{1}=K_{1}$, $Q_{2}=K_{2}$, $Q_{3}$ and $K_{3}$ were given in (\ref{Q1}), (\ref{Q2}), (\ref{Q3}) and (\ref{K3}), respectively.

\newpage

\begin{figure}[ht]  
  \begin{center}
    \vbox{
      \subfigure[]{\includegraphics[width=0.18\columnwidth]{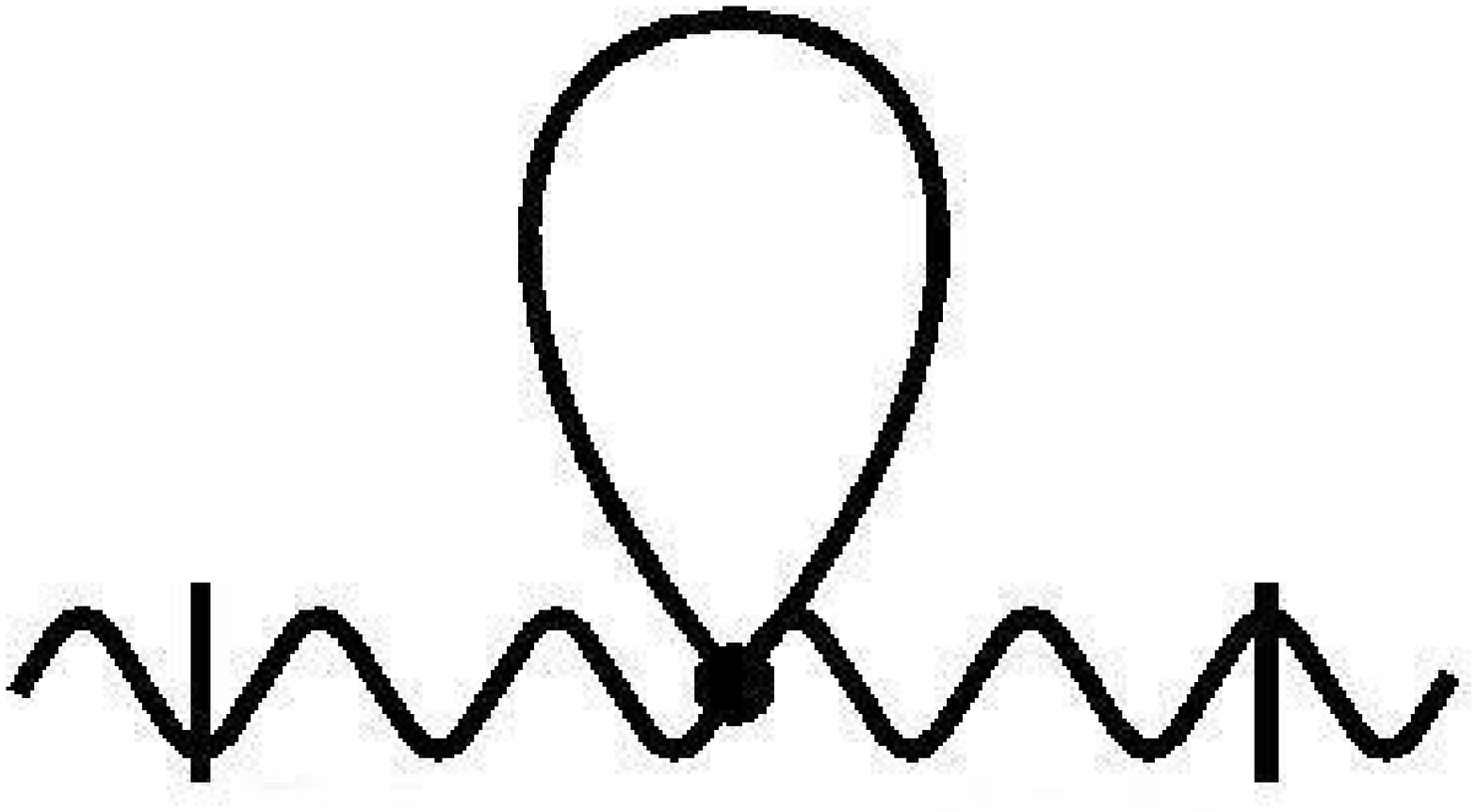}\label{fig:Figura1A}}
      \qquad
      \subfigure[]{\includegraphics[width=0.22\columnwidth]{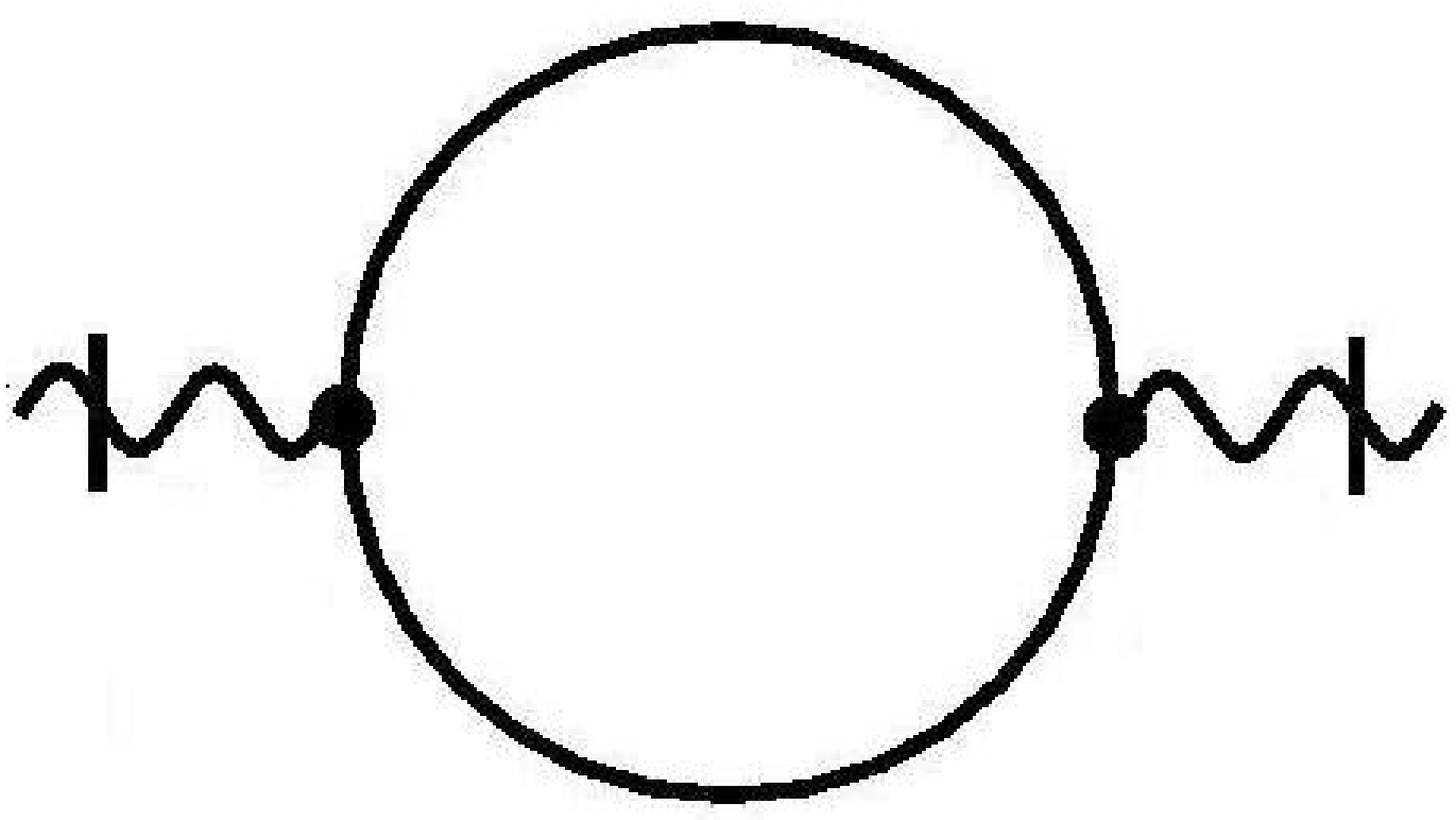}\label{fig:Figura1B}}
    }
  \end{center}
  \caption{\label{fig:Figura1} Graphs contributing to the polarization tensor (the continuous and wavy lines  represent the scalar and gauge field propagators): (a) the tadpole graph and (b) the fish graph. }
  
  \begin{center}
    \vbox{
      \subfigure[]{\includegraphics[width=0.18\columnwidth]{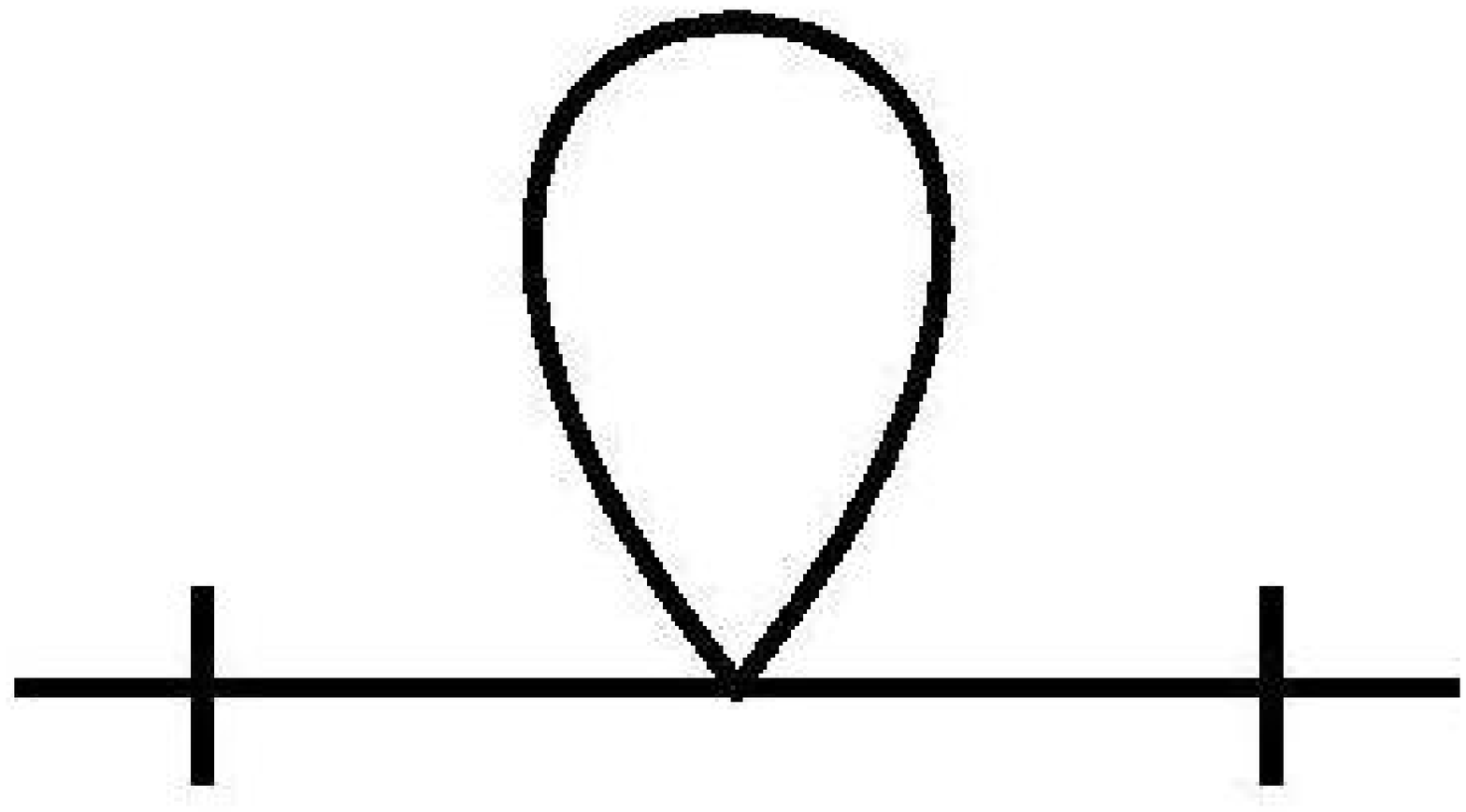}\label{fig:Figura2A}}
      \qquad
      \subfigure[]{\includegraphics[width=0.18\columnwidth]{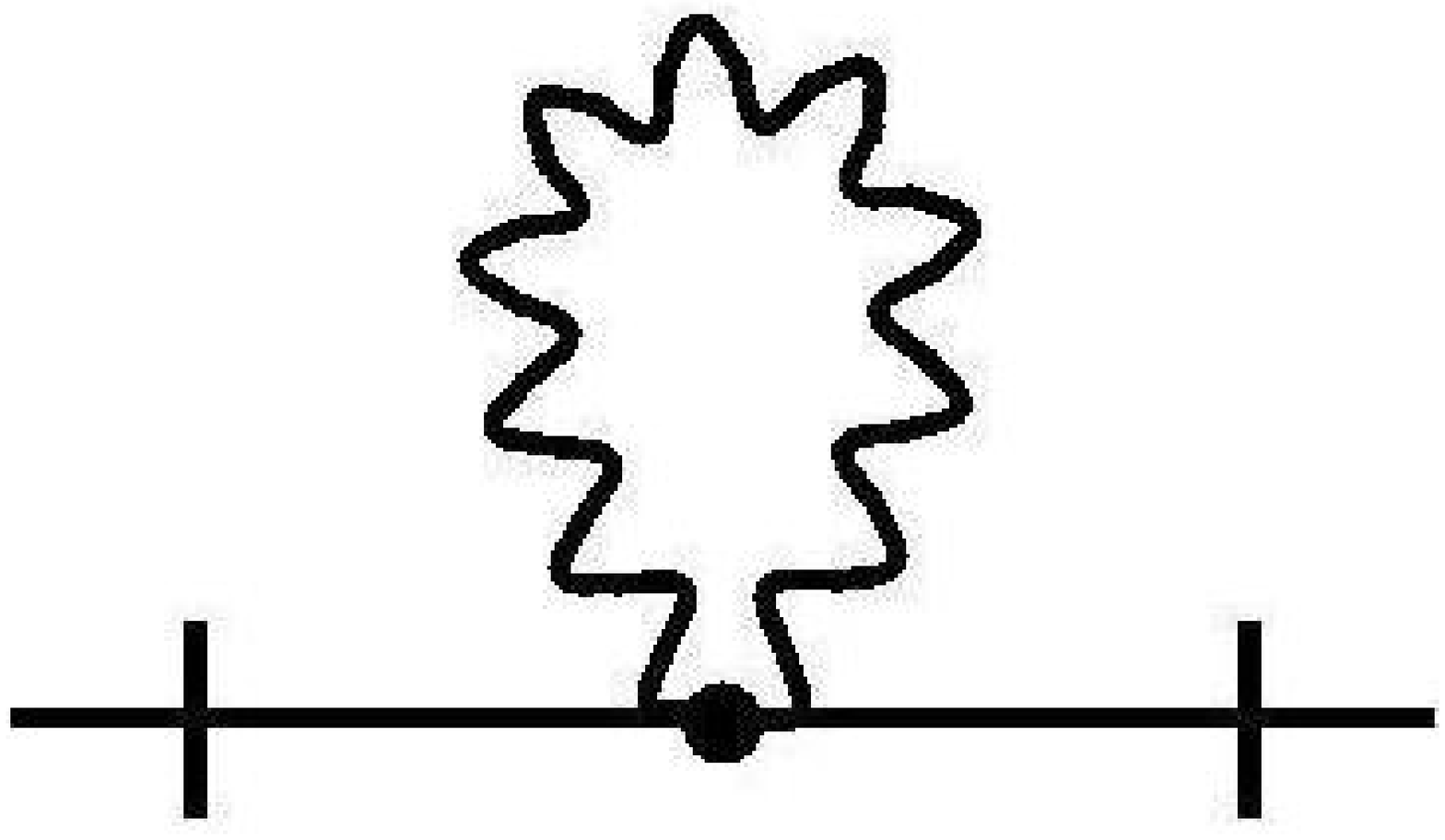}\label{fig:Figura2B}}
      \qquad
      \subfigure[]{\includegraphics[width=0.22\columnwidth]{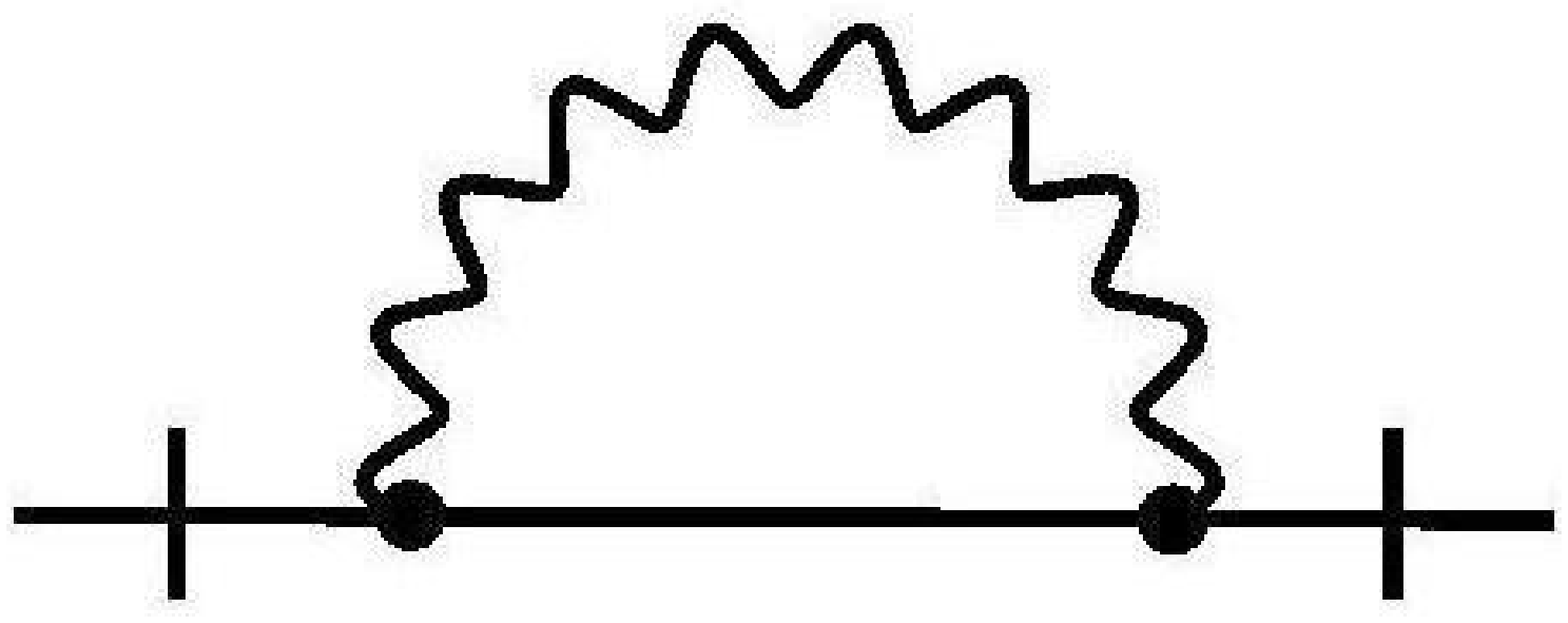}\label{fig:Figura2C}}
    }
  \end{center}
  \caption{\label{fig:Figura2}  Radiative corrections for the scalar matter field: (a) a tadpole graph with an internal scalar line,
(b) a tadpole graph with internal gauge field line and (b) a graph with two vertices. }

  \begin{center}
    \vbox{
      \subfigure[]{\includegraphics[width=0.22\columnwidth]{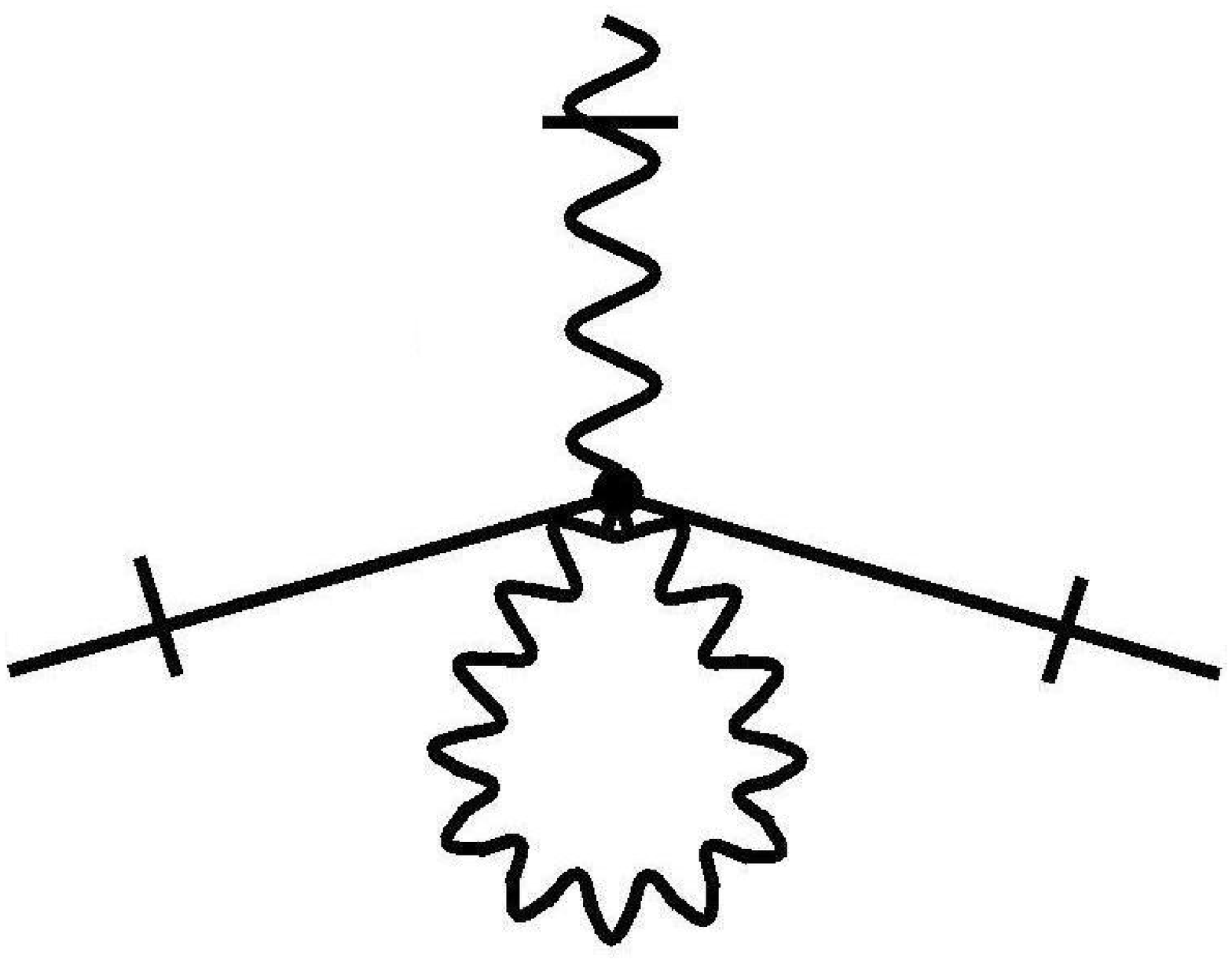}\label{fig:Figura3A}}
      \qquad \qquad
      \subfigure[]{\includegraphics[width=0.22\columnwidth]{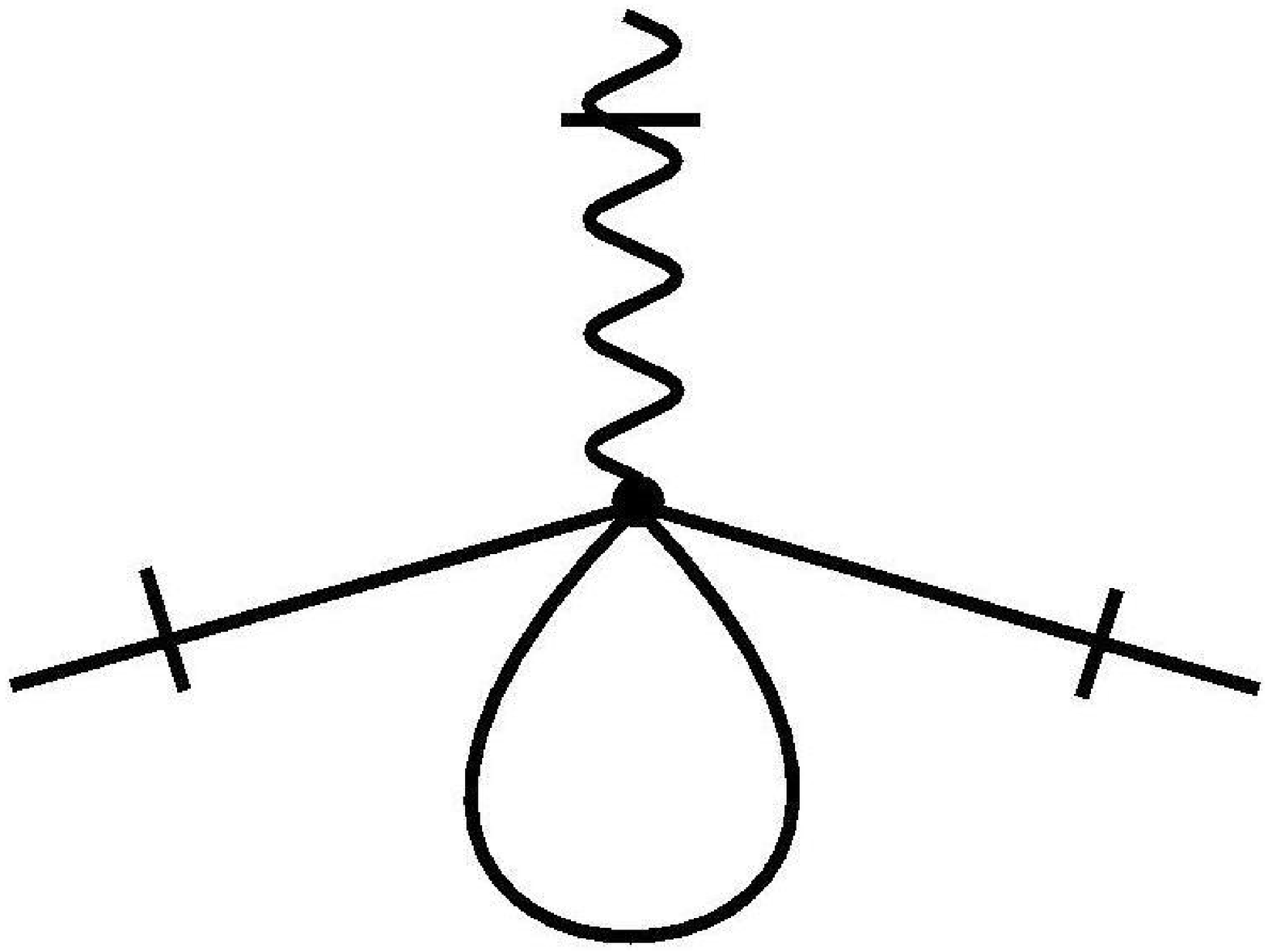}\label{fig:Figura3B}}
      \qquad \qquad
      \subfigure[]{\includegraphics[width=0.22\columnwidth]{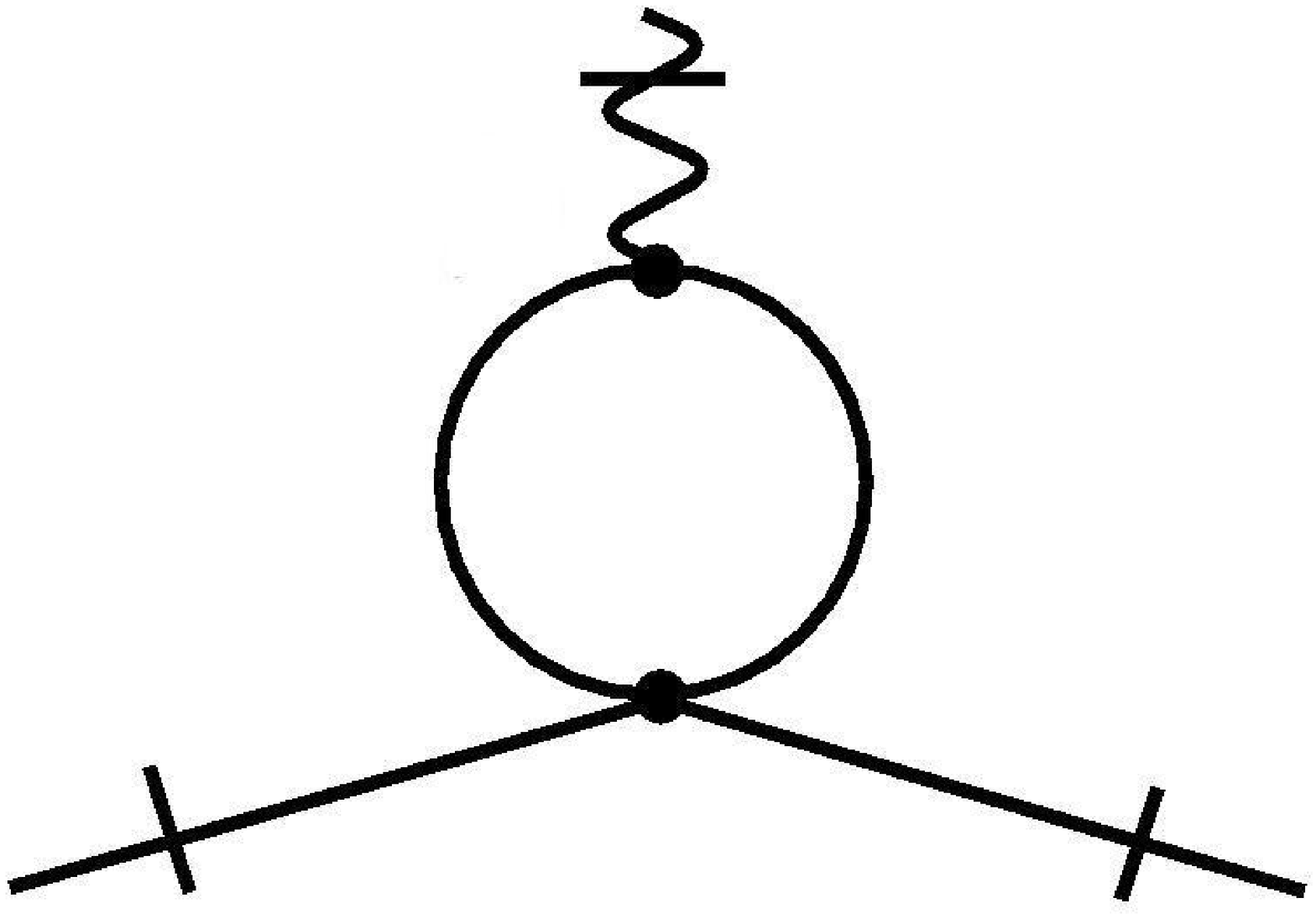}\label{fig:Figura3C}}
      \\
      \subfigure[]{\includegraphics[width=0.22\columnwidth]{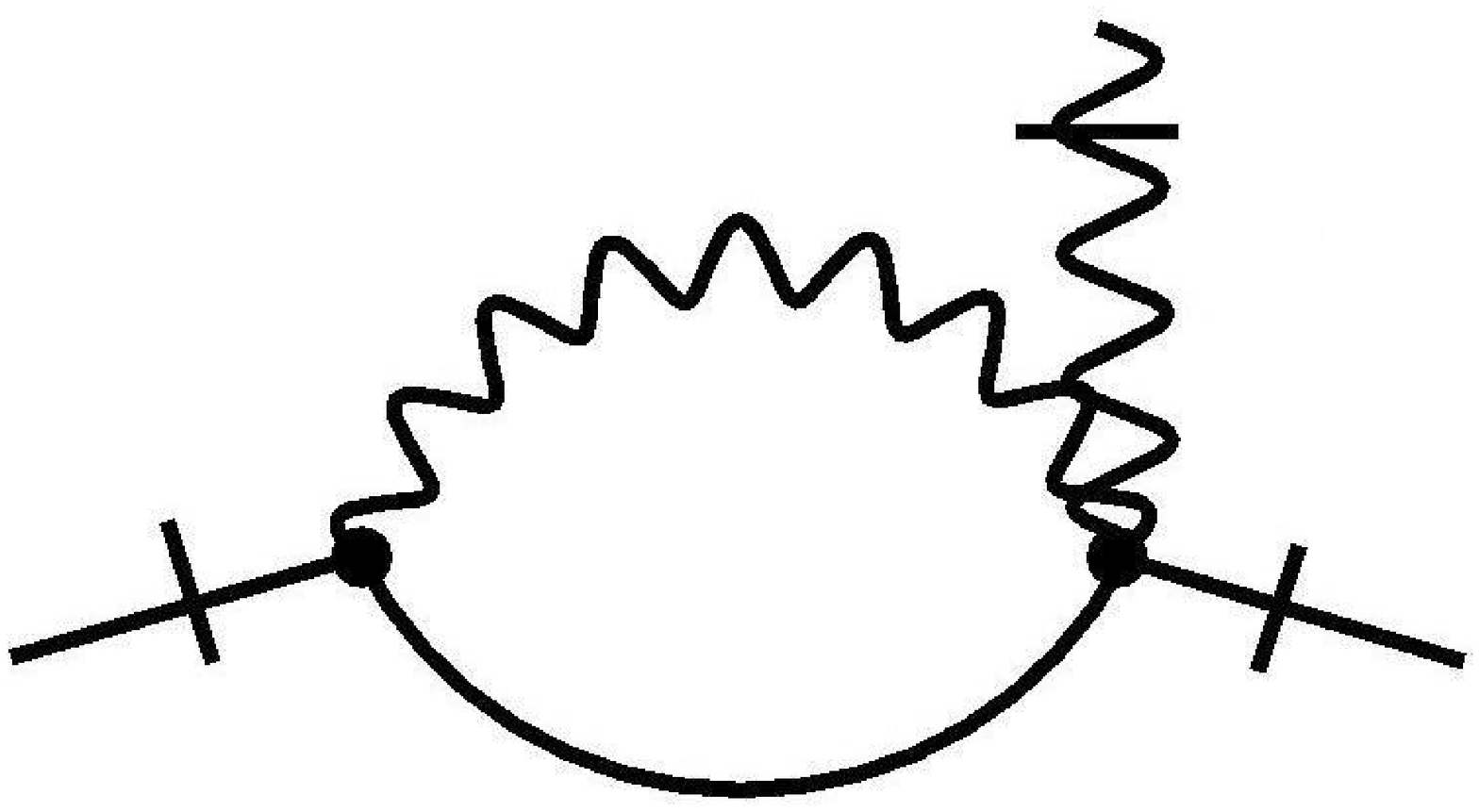}\label{fig:Figura3D}}
      \qquad \qquad
      \subfigure[]{\includegraphics[width=0.22\columnwidth]{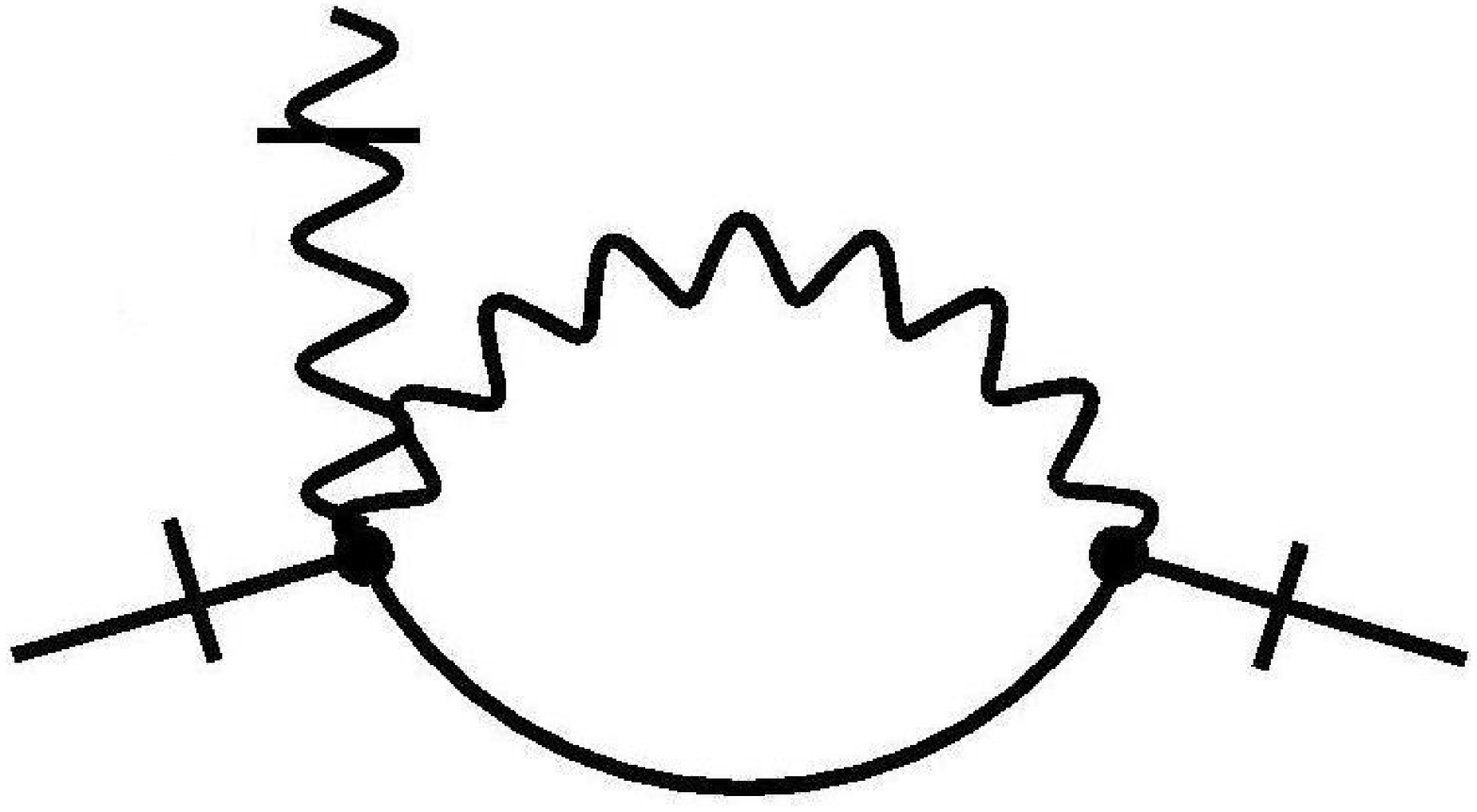}\label{fig:Figura3E}}
      \qquad \qquad
      \subfigure[]{\includegraphics[width=0.22\columnwidth]{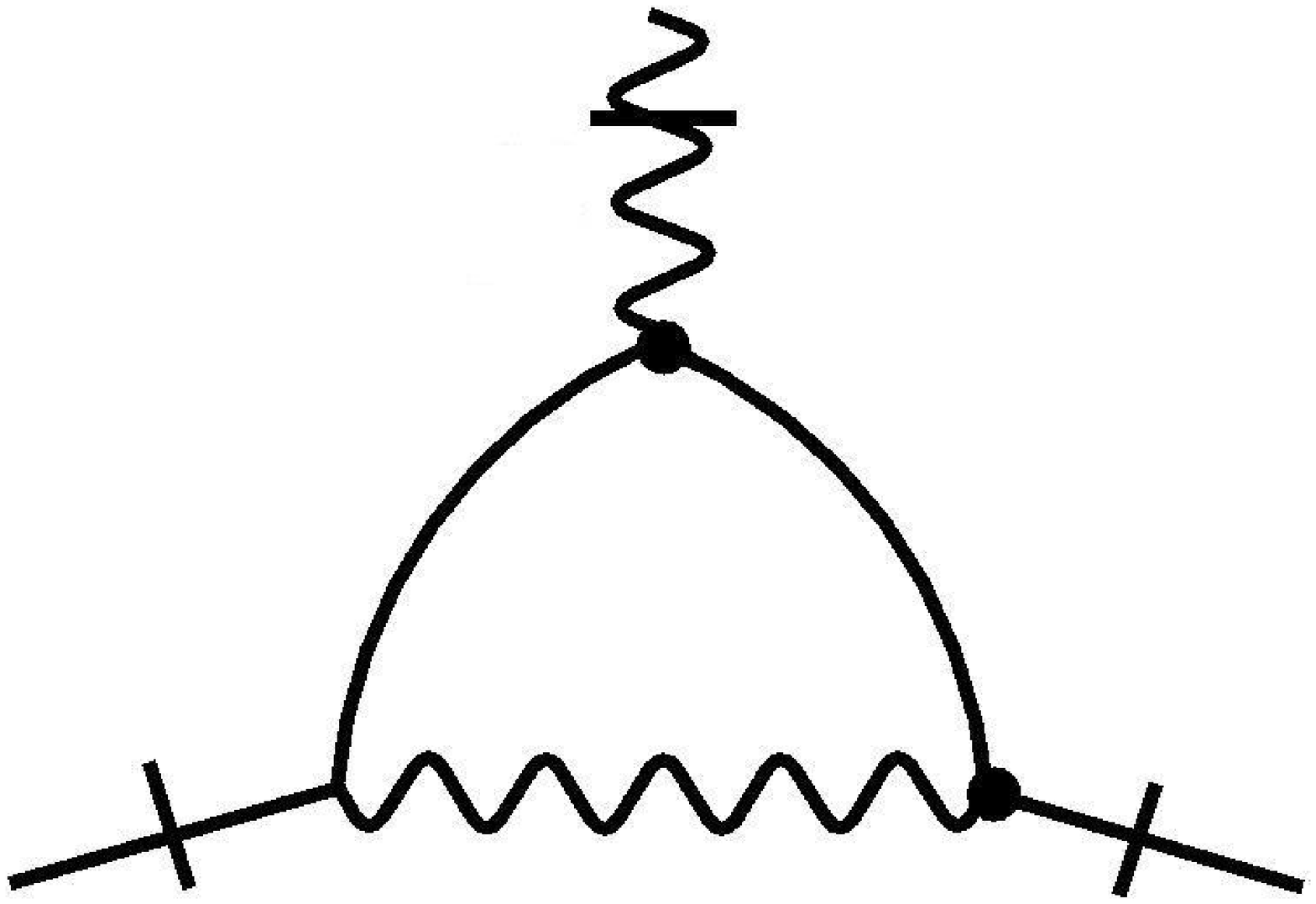}\label{fig:Figura3F}}
    }
  \end{center}
  \caption{\label{fig:Figura3}General aspect of graphs contributing to the three point vertex function.}

\end{figure} 

\end{document}